# The Structure and Dynamics of Co-Citation Clusters: A Multiple-Perspective Co-Citation Analysis

**Chaomei Chen[1,3], Fidelia Ibekwe-SanJuan[2], Jianhua Hou[3]**
[1]College of Information Science and Technology, Drexel University, 3141 Chestnut Street, Philadelphia 19104-2875, U.S.A. E-mail: chaomei.chen@drexel.edu
[2]ELICO, University of Lyon-3. 4 cours Albert Thomas, 69008 Lyon, France. E-mail: fidelia.ibekwe-sanjuan@univ-lyon3.fr
[3]WISELAB, Dalian University of Technology, Dalian, China. Email: hqzhixing@gmail.com

## Abstract

**A multiple-perspective co-citation analysis method is introduced for characterizing and interpreting the structure and dynamics of co-citation clusters. The method facilitates analytic and sense making tasks by integrating network visualization, spectral clustering, automatic cluster labeling, and text summarization. Co-citation networks are decomposed into co-citation clusters. The interpretation of these clusters is augmented by automatic cluster labeling and summarization. The method focuses on the interrelations between a co-citation cluster's members and their citers. The generic method is applied to a three-part analysis of the field of Information Science as defined by 12 journals published between 1996 and 2008: 1) a comparative author co-citation analysis (ACA), 2) a progressive ACA of a time series of co-citation networks, and 3) a progressive document co-citation analysis (DCA). Results show that the multiple-perspective method increases the interpretability and accountability of both ACA and DCA networks.**

## Introduction

Identifying the nature of specialties in a scientific field is a fundamental challenge for information science (Morris & Van der Veer Martens, 2008; Tabah, 1999). The growing interest in mapping and visualizing the structure and dynamics of specialties is due to a number of reasons:

1) the widely accessible bibliographic data sources such as the *Web of Science*, *Scopus*, and *Google Scholar* (Bar-Ilan, 2008; Meho & Yang, 2007) as well as domain-specific repositories such as *ADS*[1] and *arXiv*[2];
2) freely available computer programs and web-based general-purpose visualization and analysis tools such as *ManyEyes*[3] and *Pajek*[4] (Batagelj & Mrvar, 1998), special-purpose citation analysis tools such as *CiteSpace*[5] (Chen, 2004; Chen, 2006), social network analysis such as *UCINET*[6];
3) the intensified challenges for digesting the vast volume of data from multiple sources (e.g., e-Science, Digging into Data[7], cyber-enabled discovery, SciSIP) (Lane, 2009).

---

[1] http://www.adsabs.harvard.edu/
[2] http://arxiv.org/
[3] http://manyeyes.alphaworks.ibm.com/
[4] http://vlado.fmf.uni-lj.si/pub/networks/pajek/
[5] http://cluster.cis.drexel.edu/~cchen/citespace/
[6] http://www.analytictech.com/ucinet6/ucinet.htm
[7] http://www.diggingintodata.org/

Co-citation studies are among the most common used methods in quantitative studies of science, especially including *Author Co-citation Analysis* (ACA) (Chen, 1999; Leydesdorff, 2005; White & McCain, 1998; Zhao & Strotmann, 2008b) and *Document Co-citation Analysis* (DCA) (Chen, 2004; Chen, 2006; Chen, Song, Yuan, & Zhang, 2008; Small & Greenlee, 1986; Small & Sweeney, 1985; Small, Sweeney, & Greenlee, 1985). Co-citation relations serve a fundamental grouping mechanism in co-citation studies. Researchers typically identify specialties in terms of aggregations of co-cited individual items. The ultimate goal is to obtain insights into emergent patterns. However, interpreting patterns identified in such studies remains a major bottleneck of the entire analytical process because essential tasks for interpretation such as categorization, summarization, synthesis, and integration are not only cognitively demanding but also inadequately supported. For instance, once co-citation clusters are identified, assigning the most meaningful labels for these clusters is currently a challenging task because any representative labels of clusters must characterize not only what clusters appear to represent, but also the salient and unique reasons for their formation.

In this article, we introduce a generic multiple-perspective method for both ACA and DCA studies. Our goal is to reduce the complexity of the sense-making and synthesizing process and to provide an enriched set of cues to facilitate a variety of common tasks in interpretation. The new procedure reduces analysts' cognitive burden by automatically characterizing the nature of a co-citation cluster in terms of 1) salient noun phrases extracted from titles, abstracts, and index terms of citing articles and 2) representative sentences as summarizations of clusters. The traditional co-citation analysis typically focuses on cited members of clusters as a primary source of evidence for interpretation. Focusing on citers can improve our understanding of the nature of a research front and its intellectual base.

Our work contributes to quantitative studies of science in two ways: 1) developing a generic and consistent procedure for both ACA and DCA studies, and 2) increasing the interpretability and accountability of co-citation analysis by shifting a considerable amount of burden from analysts to automatically generated structural and content cues.

The rest of the article is organized as follows. First, we outline existing studies of scientific knowledge domains and the field of information science in particular. We focus on their methodological strengths and weaknesses and major specialties identified in the field of information science. Next, we present a new, multiple-perspective, co-citation analysis methodology, including basic metrics and various algorithmic components. We then demonstrate the use of the method through a comparative ACA with reference to an earlier ACA, an ACA, and a DCA of the information science field, enlisting the support of sense-making cues produced by spectral clustering, automatic cluster labeling, abstract-based summarization, and interactive visualization. Finally, we discuss the strengths and limitations of the new approach and directions for future research and practice.

## Co-Citation Analyses

The number of co-citation studies is increasing not only in the field of information science but also in an increasingly larger number of fields (See Figure 1). There are many self-reflective studies of the field of information science by information scientists. In this section, we outline existing studies of information science, in particular, two major types of co-citation analyses and their major findings about the structure of the field.

### *Author Co-citation Analysis (ACA) of Information Science*

*Author Co-citation Analysis* (ACA) aims to identify underlying specialties in a field in terms of groups of authors who were cited together in relevant literature. An ACA study typically focuses on a network of cited authors connected by co-citation links.

Traditionally, ACA studies are limited to the first author of a cited reference only, primarily due to the lack of required data to perform all-author analysis. More recent studies have confirmed that all-author co-citation patterns reveal stronger groupings than first-author only patterns (Schneider, Larsen, & Ingwersen, 2009). In our study, first-author ACAs are used due to the data constraints.

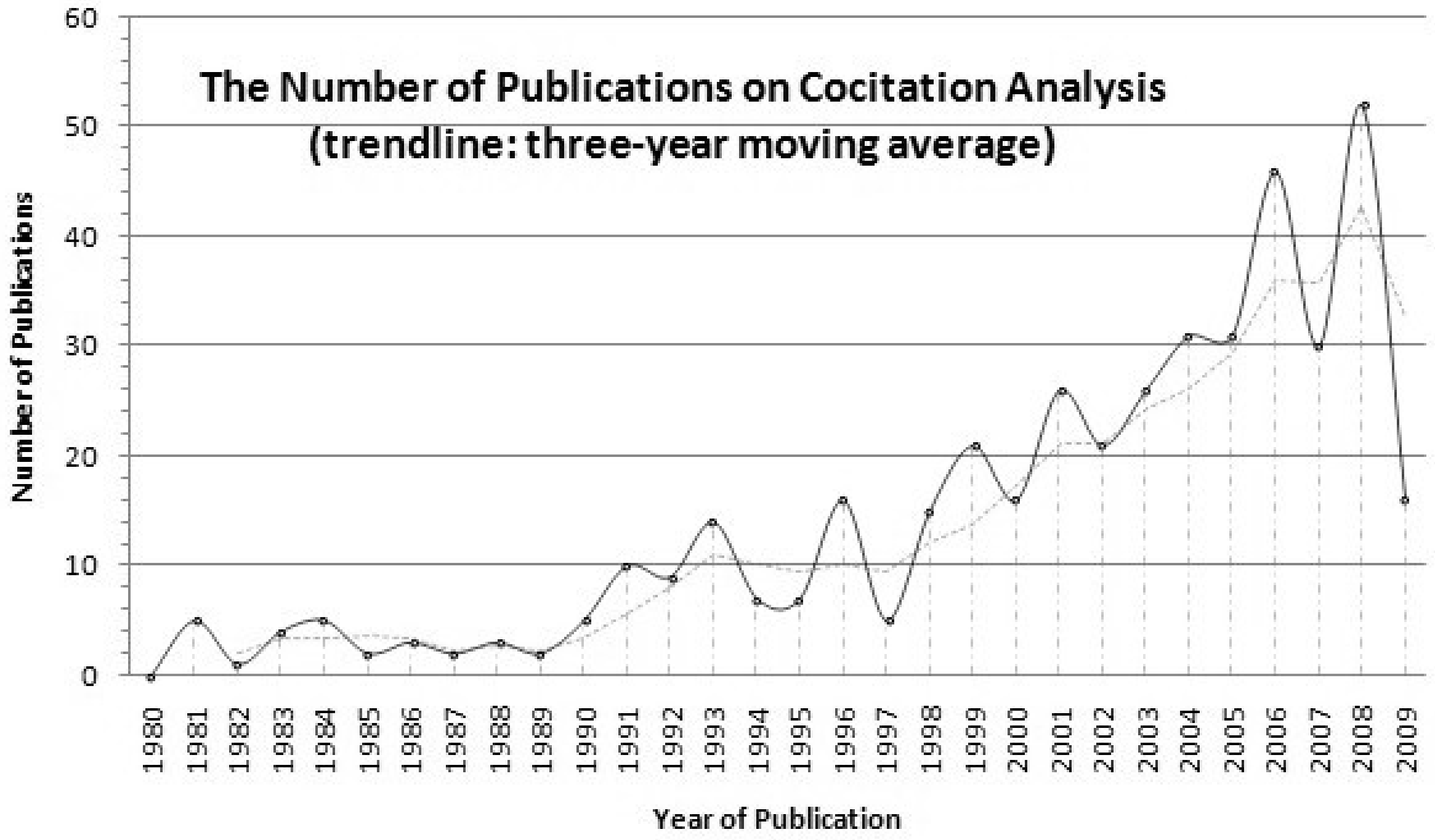

**Figure 1. The growing number of co-citation analysis publications according to a topic search query of "*co-citation* OR *cocitation*" in the Web of Science.**

**ACA of Information Science (1972-1995) by White & McCain (1998)**

This widely known study has inspired many subsequent ACA studies. It presented a comprehensive view of information science based on 12 journals in library and information science across a 24-year span (1972-1995). It analyzed co-citation patterns of 120 most cited authors with factor analysis and multidimensional scaling. The authors drew upon their extensive knowledge of the field and offered an insightful interpretation of 12 specialties identified in terms of 12 factors. The most well-known finding of the study is that information science at the time consisted of two essentially independent camps, namely the information retrieval camp and the literature camp, including citation analysis, bibliometrics, and scientometrics. This study has become an exemplar of ACA studies in terms of its methodology and its account of information science as a discipline.

**ACA and ABCA of Information Science (1996-2005) by Zhao & Strotmann (2008a, 2008b)**

Zhao and Strotmann (2008a, 2008b) followed up White and McCain's study using the same set of 12 journals and the same number of 120 cited authors in an updated time frame of 1996-2005. Zhao and Strotmann (2008a) studied the intellectual base, defined by co-cited authors (ACA), and the patterns of citing authors called *author-bibliographic coupling analysis* (ABCA).

Zhao and Strotmann (2008b) found five major specialties and manually labeled them as *user studies*, *citation analysis*, *experimental retrieval*, *webometrics*, and *visualization of knowledge domains*. In contrast to the findings of (White & McCain, 1998), *experimental retrieval* and *citation analysis* retained their fundamental roles in the field, and the *user studies* specialty became the largest specialty. *Webometrics* and *visualization of knowledge domains* appeared to make connections between the retrieval camp and the citation analysis camp.

In their second study, Zhao and Strotmann (2008a) adapted the *bibliographic coupling* method introduced by Kessler (1963) and analyzed co-citing authors (ABCA). They compared the intellectual base (ACA) and research fronts (co-citing authors) during the period of 1996-2005. Again, the study was limited to the topmost 120 citing authors using factor analysis and principle component analysis (PCA) to identify specialties.

Their analysis of current citing authors found that the global two-camp structure was still evident. They also found that the literatures camp, namely *citation* and *bibliometrics*, "remained remarkably stable and unperturbed by the Web revolution", and that *webometrics* became an important component in this camp. Furthermore, *webometrics* was more active during 1996-2000 than 2001-2005.

*Document Co-citation Analysis (DCA)*

Document Co-citation Analysis (DCA) studies a network of co-cited references (Small, 1980, 2003). The fundamental assumption is that co-citation clusters reveal underlying intellectual structures. The notion that cited documents are in effect concept symbols was introduced by Small (1978). He found a high degree of uniformity in how specific concepts and specific references to documents (cited documents) were associated in chemistry literature. These cited documents serve as symbols for scientific ideas, methods, and experiments. This idea is further extended to clusters of noun phrases extracted from citation contexts of cited documents (Schneider, 2006). From the concept symbol perspective, the study of a co-citation network focuses on interpreting the nature of a cluster of cited documents and interrelationships between clusters.

DCA is expected to reveal more specific patterns than ACA because cited references in DCA carry more specific information than cited authors in ACA. The nature of the use of a cited reference in a DCA network should be relatively easier to identify and less ambiguous to interpret than the nature of the use of a cited author in an ACA network. We expect that both DCA and ACA networks would lead us to unique insights and it will be valuable to establish a comprehensive methodology that can be consistently applied to both types of analysis so that one can compare and cross reference the results.

**DCA of Library and Information Science (1990-2004) by Aström (2007)**

In contrast to ACA studies, there are much fewer DCA studies in the literature. A DCA by Aström (2007) studied papers published between 1990 and 2004 in 21 library and information science journals. Results were depicted in multidimensional scaling (MDS) maps. Aström's study also identified the two-camp structure found by (White & McCain, 1998). On the other hand, Aström found an *information seeking and retrieval* camp, instead of the *information retrieval* camp as in (White and McCain, 1998).

*Determining the Nature of Specialties: A Bottleneck*

An increasing number of analytical tasks in ACA and DCA are now facilitated by computer programs. However, researchers and analysts still have to deal with a large amount of diverse and complex information. We address some of the current challenges and how we propose to facilitate the process.

Both ACA and DCA studies focus on structural patterns of co-cited authors or references. It is a common practice in ACA studies to identify research specialties in terms of co-citation clusters or multivariate factors and then interpret the nature of these specialties. While manually labeling a co-citation cluster can be a very rewarding process of learning about the underlying specialty and result in insightful and easy to understand labels, it requires a substantial level of domain knowledge and it tends to be time consuming and cognitively demanding because of the synthetic work required over a diverse range of individual publications. In addition, it is hard for others to follow the heuristics and criteria used in the original analytic reasoning process.

Traditionally, researchers often identify the nature of a co-citation cluster based on common themes among its members. For example, White and McCain (1998) tapped on their decades of domain knowledge and identified what members of specialties have in common. Similarly, Zhao and Strotmann (2008a, 2008b) also focused on common areas shared by member authors of each factor. The emphasis on common areas is a practical strategy; otherwise, comprehensively identifying the nature of a specialty can be too complex to handle manually. In addition to identify common characteristics, co-citation analysts need to synthesize a diverse range of potentially conflicting information, aggregate and categorize lower level evidence to arrive at a coherent interpretation at a higher level abstraction. One can leverage algorithmically generated semantic cues so as to facilitate these cognitively demanding tasks and reduce potential biases and uncertainties.

Algorithmically labeling co-citation clusters is a crucial step towards improving the overall efficiency of citation analysis. The *Institute for Scientific Information* (ISI) and others pioneered various approaches. A fundamental challenge for automatic labeling is due to the practically limited vocabulary that one can possibly extract from a given data source because the best possible candidates may never appear in the data per se. A promising strategy is to expand the vocabulary pool by tapping into multiple data sources such as the *Wikipedia* (Carmel, Roitman, & Zwerdling, 2009). Similarly, Zuccala (2006) selected cluster labels from classification codes assigned to papers written by authors in a cluster. Our goal is to facilitate an otherwise entirely manual labeling process so that analysts may evaluate and choose algorithmically generated terms. It is our intention that human analysts will still maintain a full control of the interpretation and sense making stages of the citation analysis.

### *Identifying Specialties in Information Science*

How is the field of information science itself organized? What are the major components in its structure? How many specialties are there? These questions have been addressed in the literature. Now we are seeking additional insights by using a multiple-perspective approach.

Many researchers have studied the structural and dynamic properties of specialties in information science in terms of clusters, multivariate factors, and principle components (Morris & Van der Veer Martens, 2008; Persson, 1994; Tabah, 1999; White & Griffith, 1982). A recent study of information science (Ibekwe-SanJuan, 2009) mapped the structure of information science at the term level using a text analysis system *TermWatch* and a network visualization system *Pajek*, but it did not address structural patterns of cited references. Researchers also studied the structure of information science qualitatively, especially with direct inputs from domain experts. For example, Zins conducted a *Critical Delphi* study of information science, involving 57 leading information scientists from 16 countries (Zins, 2007a, 2007b, 2007c, 2007d). Our present study focuses on quantitative approaches.

Janssens, Leta, Glänzel, and De Moor (2006) studied the full-text of 938 publications in five library and information science journals with Latent Semantic Analysis (LSA) (Deerwester, Dumais, Landauer, Furnas, & Harshman, 1990) and agglomerative clustering. They found an optimal 6-cluster solution in terms of a local maximum of the mean silhouette coefficients (Rousseeuw, 1987) and a stability diagram (Ben-Hur, Elisseeff, & Guyon, 2002). Their clusters were labeled with single-word terms selected by tf*idf (p. 1625), which are not as informative as multiword terms for cluster labels.

Klavans, Persson, and Boyack (2009) recently raised the question of the true number of specialties in information science. They suspected that the number is much more than the 11 or 12 as reported in ACA studies such as (White & McCain, 1998) and (Zhao & Strotmann, 2008a, 2008b), but significantly fewer than the 72 reported in their own study, which is also based on the 12 journals between 2001 and 2005.

## Method

Building on the work of these relevant studies, we introduce a multiple-perspective co-citation analysis method. The term *multiple-perspective* refers to the analysis of structural, temporal, and semantic patterns as well as the use of both citing and cited items for interpreting the nature of co-citation clusters. We focus on ACA and DCA in this article. In this section, we first set the context of our work with reference to the traditional procedure and introduce several citation-related and structure-related metrics for subsequent discussions. Then we explain three components of the new procedure, namely, clustering, labeling, and sentence selection. We briefly introduce *CiteSpace*, the system we use to perform the study.

### *Data*

We applied the new method to a domain analysis of information science as defined by the same 12 journals used by White and McCain (1998) and Zhao and Strotmann (2008a) but over a longer time span (1996-2008). Since the initial selection of the 12 journals was more than a decade ago, they may no longer be the best cohort to track the field of information science. On the other hand, from the point of view of validation and continuity with regard to previous studies, these journals are rich enough to demonstrate the multiple-perspective method.

The 12-journal *Information Science* dataset, retrieved from the *Web of Science*, contains 10,853 unique bibliographic records, written by 8,408 unique authors from 6,553 institutions and 89 countries. These articles cited 129,060 unique references for a total of 206,180 times. They cited 58,711 unique authors and 58,796 unique sources. See Table 1 for details.

**Table 1. The Information Science (1996-2008) dataset defined by 12 source journals.**

| Start | End | Citing Sources: Journal Title | Abbr. | Records | Citations (TC) | Cites per Paper |
|---|---|---|---|---|---|---|
| 1996 | 2008 | Annual Review of Information Science and Technology | ARIST | 158 | 1,584 | 10.03 |
| 1996 | 2008 | Electronic Library | EL | 1,346 | 835 | 0.62 |
| 1996 | 2008 | Information Processing & Management | IP&M | 1,037 | 6,398 | 6.17 |
| 1996 | 2008 | Information Technology and Libraries | ITL | 474 | 721 | 1.52 |
| 1996 | 2008 | Journal of Documentation | JOD | 1,057 | 3,716 | 3.52 |
| 1996 | 2008 | Journal of Information Science | JIS | 653 | 2,510 | 3.84 |
| 1996 | 2000 | Journal of the American Society for Information Science | JASIS | 851 | 8,211 | 9.65 |
| 2001 | 2008 | Journal of the American Society for Information Science and Technology | JASIST | 1,410 | 6,857 | 4.86 |
| 1996 | 2008 | Library & Information Science Research | LISR | 565 | 1,462 | 2.59 |
| 1996 | 2008 | Library Resources & Technical Services | LRTS | 542 | 579 | 1.07 |
| 1996 | 1996 | PROGRAM-Automated Library and Information Systems | P-ALIS | 124 | 41 | 0.33 |
| 1996 | 2008 | PROGRAM-Electronic Library and Information Systems | P-ELIS | 1,215 | 376 | 0.31 |
| 1996 | 2008 | Scientometrics | SCM | 1,421 | 8,447 | 5.94 |

*Extending the Traditional Procedure*

The primary goal of co-citation analysis is to identify the intellectual structure of a scientific knowledge domain in terms of the groupings formed by accumulated co-citation trails in scientific literature. The traditional procedure of co-citation analysis for both DCA and ACA consists of the following steps:

1. Retrieve citation data from sources such as the *Science Citation Index* (SCI), *Social Science Citation Index* (SSCI), *Scopus*, and *Google Scholar*.
2. Construct a matrix of co-cited references (DCA) or authors (ACA).
3. Represent the co-citation matrix as a node-and-link graph or as a multidimensional scaling (MDS) configuration with possible link pruning using Pathfinder network scaling or minimum spanning tree algorithms.
4. Identify specialties in terms of co-citation clusters, multivariate factors, principle components, or dimensions of a latent semantic space using a variety of algorithms for clustering, community-finding, factor analysis, principle component analysis, or latent semantic indexing.

5. Interpret the nature of co-citation clusters.

The interpretation step is the weakest link. It is time-consuming and cognitively demanding, requiring a substantial level of domain knowledge and synthesizing skills. In addition, much of attention routinely focuses on co-citation clusters per se, but the role of citing articles that are responsible for the formation of such co-citation clusters may not be always investigated as an integral part of a specialty. While the focus on the target of citation may reveal seminal members of a specialty, it does not necessarily reflect the dynamics of the specialty in terms of the impact on its citers. Researchers have used citing information to summarize the essence of a co-citation cluster. For example, Small (1986) introduced a method for generating specialty narratives by walking through a co-citation network and selecting passages that cite the core documents in a co-citation cluster. There are also other studies that take citing information into account (Schneider, 2009).

Given the diversity and complexity of relationships between citers and cited entities (Cronin, 1981), synthesizing the nature of a co-citation cluster is cognitively too demanding for analysts to handle manually. The lack of algorithmic support for these tasks forces analysts to rely on their own domain knowledge and their experience. It makes it hard to differentiate evidence-based findings from heuristics and speculations. Such ambiguity may hinder subsequent evaluation and scholarly communication of research findings. These problems motivate us to develop a multiple-perspective method to improve the robustness of the traditional procedure.

Our new method extends and enhances traditional co-citation methods in two ways: 1) by integrating structural and content analysis components sequentially into the new procedure and 2) by facilitating analytic tasks and interpretation with automatic cluster labeling and summarization functions. The new procedure is highlighted in yellow in Figure 2, including clustering, automatic labeling, summarization, and latent semantic models of the citing space (Deerwester et al., 1990).

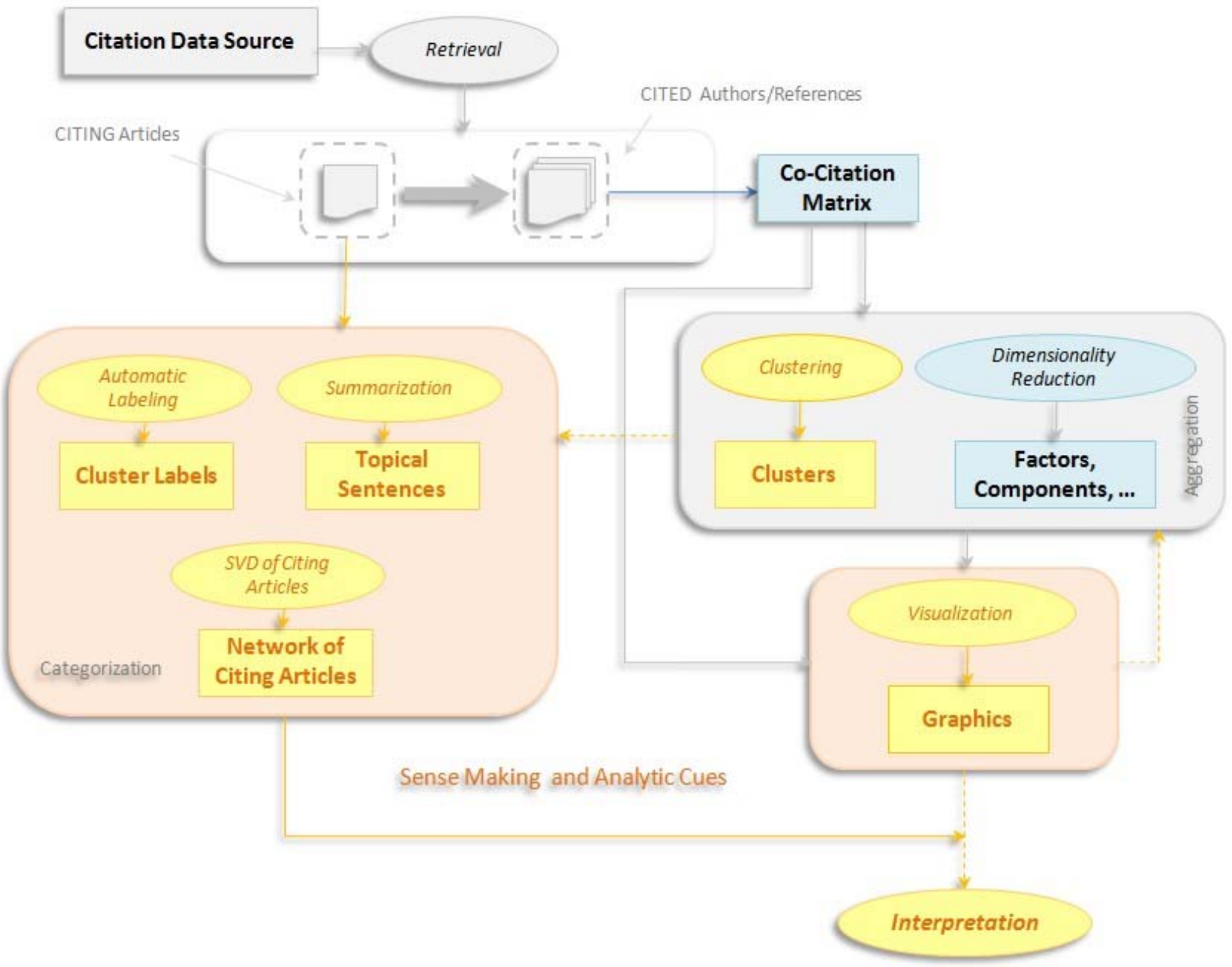

**Figure 2. New functions facilitate and enhance the interpretation of co-citation clusters.**

*Metrics*

Our new procedure adopts several structural and temporal metrics of co-citation networks and subsequently generated clusters. Structural metrics include betweenness centrality, modularity, and silhouette. Temporal and hybrid metrics include citation burstness and novelty.

The betweenness centrality metric is defined for each node in a network. It measure the extent to which the node is in the middle of a path that connects other nodes in the network (Brandes, 2001; Freeman, 1977). High betweenness centrality values identify potentially revolutionary scientific publications (Chen, 2005) as well as gatekeepers in social networks. For example, if a node provides the only connection between two large but otherwise unrelated clusters, then this author would have a very high value of betweenness centrality. Recently, power centrality introduced by Bonacich (1987) is also drawing a lot of attention in dealing with networks in which someone's power depends on the power of those (s)he is socially related to, for example, in (Kiss & Bichler, 2008).

In the context of this study, the modularity Q measures the extent to which a network can be divided into independent blocks, i.e. modules (Newman, 2006; Shibata, Kajikawa, Taked, & Matsushima, 2008). The modularity score ranges from 0 to 1. A low modularity suggests a network that cannot be reduced to clusters with clear boundaries, whereas a high modularity may imply a well-structured network. On the other hand, networks with modularity scores of 1 or very close to 1 may turn out to be some trivial special cases where individual components are simply isolated from one another. Since the modularity is defined for any network, one may compare different networks in terms of their modularity, for example, between ACA and DCA networks.

The silhouette metric (Rousseeuw, 1987) is useful in estimating the uncertainty involved in identifying the nature of a cluster. The silhouette value of a cluster, ranging from -1 to 1, indicates the uncertainty that one needs to take into account when interpreting the nature of the cluster. The value of 1 represents a perfect separation from other clusters. In this study, we expect that cluster labeling or other aggregation tasks will become more straightforward for clusters with the silhouette value in the range of 0.7~0.9 or higher.

Burst detection determines whether a given frequency function has statistically significant fluctuations during a short time interval within the overall time period. It is valuable for citation analysts to detect whether and when the citation count of a particular reference has surged. For example, after the September 11 terrorist attacks, citations to earlier studies of Oklahoma City Bombing were increased abruptly (Chen, 2006). It can be also used to detect whether a particular connection has been significantly strengthened within a short period of time (Kumar, Novak, Raghavan, & Tomkins, 2003). We adopt the burst detection algorithm introduced in (Kleinberg, 2002).

Sigma ($\sum$) is introduced in (Chen, Chen, Horowitz, Hou, Liu, & Pellegrino, 2009a) as a measure of scientific novelty. It identifies scientific publications that are likely to represent novel ideas according to two criteria of transformative discovery. As demonstrated in case studies (Chen et al., 2009a), Nobel Prize and other award winning research tends to have highest values of this measure. In this study, Sigma is defined as $(centrality+1)^{burstness}$ such that the brokerage mechanism plays more prominent role than the rate of recognition by peers.

*Clustering*

We adopt a hard clustering approach such that a co-citation network is partitioned to a number of non-overlapping clusters. It is more efficient to use non-overlapping clusters than overlapping ones to differentiate the nature of different co-citation clusters, although it is conceivable to derive a soft clustering version of this particular component. Resultant clusters are subsequently labeled and summarized.

In this article, co-citation similarities between items *i* and *j* are measured in terms of cosine coefficients. If *A* is the set of papers that cites *i* and *B* is the set of papers that cite *j*, then $w_{ij} = \frac{|A \cap B|}{\sqrt{|A| \times |B|}}$, where $|A|$ and $|B|$ are the citation counts of *i* and *j*, respectively; and $|A \cap B|$ is the co-citation count, i.e. the number of times they are cited together. Alternative similarity measures are also available. For example, Small (1973) used $w_{ij} = \frac{|A \cap B|}{|A \cup B|}$, which is known as the Jaccard index (Jaccard, 1901).

A good partition of a network would group strongly connected nodes together and assign loosely connected ones to different clusters. This idea can be formulated as an optimization problem in terms of a cut function defined over a partition of a network. Technical details are given in relevant literature (Luxburg, 2006; Ng, Jordan, & Weiss, 2002; Shi & Malik, 2000). A *partition* of a network *G* is defined by a set of sub-graphs $\{G_k\}$ such that $G = \bigcup_{k=1}^{K} G_k$ and $G_i \cap G_j = \emptyset$, for all $i \neq j$. Given sub-graphs *A* and *B*, a *cut function* is defined as follows: $cut(A, B) = \sum_{i \in A, j \in B} w_{ij}$, where $w_{ij}$'s are the cosine coefficients mentioned above. The criterion that items in the same cluster should have strong connections can be optimized by maximizing $\sum_{k=1}^{K} cut(G_{k,} G_k)$. The criterion that items between different clusters should be only weakly connected can be optimized by minimizing $\sum_{k=1}^{K} cut(G_k, G - G_k)$. In this study, the cut function is normalized by $\sum_{k=1}^{K} \frac{cut(G_k, G - G_k)}{vol(G_k)}$ to achieve more balanced partitions, where $vol(G_k)$ is the sum of the weights of links in $G_k$, i.e. $vol(G_k) = \sum_{i \in G_k} \sum_j w_{ij}$ (Shi & Malik, 2000).

Spectral clustering is an efficient and generic clustering method (Luxburg, 2006; Ng et al., 2002; Shi & Malik, 2000). It has roots in spectral graph theory. Spectral clustering algorithms identify clusters based on eigenvectors of Laplacian matrices derived from the original network. Spectral clustering has several desirable features compared to traditional algorithms such as *k*-means and single linkage (Luxburg, 2006):

1. It is more flexible and robust because it does not make any assumptions on the forms of the clusters;
2. It makes use of standard linear algebra methods to solve clustering problems; and
3. It is often more efficient than traditional clustering algorithms.

Our multiple-perspective method utilizes the same spectral clustering algorithm for both ACA and DCA studies. Despite its limitations (Luxburg, Bousquet, & Belkin, 2009), spectral clustering provides clearly defined information for subsequent automatic labeling and summarization to work with. In this study, instead of letting the analyst to specify how many clusters there should be, the number of clusters is uniformly determined by the spectral clustering algorithm based on the optimal cut described above.

*Automatic Cluster Labeling*

Candidates of cluster labels are selected from noun phrases and index terms of citing articles of each cluster. These terms are ranked by three different algorithms. In particular, noun phrases are extracted from titles and abstracts of citing articles. The three term ranking algorithms are *tf*idf* (Salton, Yang, & Wong, 1975), *log-likelihood ratio* (LLR) tests (Dunning, 1993), and *mutual information* (MI). Labels selected by *tf*idf* weighting tend to represent the most salient aspect of a cluster, whereas those chosen by log-likelihood ratio tests and mutual information tend to reflect a unique aspect of a cluster.

Garfield (1979) has discussed various challenges of computationally selecting the most meaningful terms from scientific publications for subject indexing. Indeed, the notion of citation indexing was originally proposed as an alternative strategy to deal with some of the challenges. White (2007a, 2007b) offers a new way to capture the relevance of a communication in terms of the widely known *tf*idf* formula.

*Automatic Summarization of Citers' Abstracts*

Each co-citation cluster is summarized by a list of sentences selected from the abstracts of articles that cite at least one member of the cluster. Sentences are selected based on the following principles.

A good summary should have a sufficient and balanced coverage with minimal redundant information (Sparck Jones, 1999). Automatic multi-document summarization typically constructs a short summary by selecting the most representative sentences from a set of topically related documents. Teufel and Moens (2002) have proposed an intriguing strategy for summarizing scientific articles based on the rhetorical status of statements in an article. Their strategy specifically focuses on identifying the new contribution of a source article and its connections to earlier work. Automatic summarization techniques have been applied to areas such as identifying drug interventions from MEDLINE (Fiszman, Demner-Fushman, Kilicoglu, & Rindflesch, 2009).

In this study, sentences are ranked by *Enertex* (Fernandez, SanJuan, & Torres-Moreno, 2007). Given a set *S* of *N* sentences, let *M* be the square matrix that for each pair of sentences gives the number of nominal words in common (nouns and adjectives). Sentences are ranked by an energy function *E*, which is defined for each sentence $s_i \in S$ as follows:

$$E(s_i) = \sum_{j=1}^{N} M^2{}_{ij}$$

Two sentences, $s_1$ and $s_2$, are not directly connected if they do not have overlapping words. However, they can be indirectly connected if there is a sentence $s_3$ such that $s_3$ overlaps with $s_1$ and $s_2$ respectively. The energy function *E* takes into account indirect connections between sentences. According to Fernandez et al. (2007), the energy function *E* in Enertex provides an efficient approximation of sentence ranking scores with regard to algorithms such as PageRank (Brin & Page, 1998), LexRank (Radev & Erkan, 2004), and TextRank (Mihalcea & Tarau, 2004).

In this study, summarization sentences were also ranked by two new functions *gtf* and $gft_{idf}$, which are further simplified approximations of the energy function *E*. Given sentences $s_i$ and $s_j$ in ***S*** and a word *w* in a set *W*, let $w_s$ be the frequency of word *w* in sentence *s*. $idf_w$ is the inverse document frequency of word *w* in sentences. Then, *gtf* and $gft_{idf}$ are defined as follows for each sentence $s_i$ with respect to ***N*** sentences. $gft_{idf}$ scores are *idf* weighted *gtf* scores.

$$gtf(s_i) = \sum_{w \epsilon W, j=1}^{N} w_{s_i} \cdot w_{s_j}$$

$$gtf_{idf}(s_i) = \sum_{w \epsilon W, j=1}^{N} w_{s_i} \cdot w_{s_j} \cdot idf_w{}^2$$

Computing *gtf* scores is faster than computing *E*, but bear in mind they are approximations to *TextRank* and *LexRank* ranking scores.

*CiteSpace*

The ACA and DCA studies described in this article were conducted using the *CiteSpace* system (Chen, 2004; Chen, 2006). *CiteSpace* is a freely available Java application for visualizing and analyzing emerging trends and changes in scientific literature. *CiteSpace* supports a unique type of co-citation network analysis – *progressive network analysis* – based on a time slicing strategy and then synthesizing a series of individual network snapshots defined on consecutive time slices. Progressive network analysis particularly focuses on nodes that play critical roles in the evolution of a network over time. Such critical nodes are candidates of intellectual turning points. Further technical details and case studies are available in (Chen, 2006). *CiteSpace* has been actively maintained and updated with new visual analytic features as well as theoretical developments (Chen et al., 2009a; Chen et al., 2008; Chen, Zhang, & Vogeley, 2009b). Metrics such as burstness, centrality, modularity, and silhouette metrics described above are implemented and incorporated in various visualizations in *CiteSpace*. For example, purple rims of nodes indicate the importance of nodes in terms of betweenness centrality (≥0.1). Three types of link similarity measures are currently supported: cosine, Dice, and Jaccard similarity coefficients.

In *CiteSpace* visualizations, clusters are labeled by candidate terms selected by ranking algorithms such as tf*idf, log-likelihood ratio tests, or mutual information. Cluster labels are placed at the weight centers of clusters in a cluster view or at the end of a timeline in a timeline view. The cluster view displays a network in a node-and-link diagram. The timeline view displays a network by arranging its clusters along horizontal timelines. The design of the timeline view is inspired by the *DIVA* system (Morris, DeYong, Wu, Salman, & Yemenu, 2002). While in *DIVA* the analyst needs to work out appropriate labels manually, in *CiteSpace* the selection of labels is supported by nine ranked lists of candidate labels.

In addition to sense-making cues such as clusters, labels, and summarizations, analysts may interact with various visualizations of co-citation clusters and their members. Interactive functions in *CiteSpace* correspond to three levels of units of analysis. At the network level, functions operate on networks, including global visualizations of networks: a node-and-link cluster view and a timeline view. At the cluster level, functions operate on individual clusters such as showing all the citers to a cluster or hiding a cluster. At the basic entity level, functions are restricted to individual entities, for example, showing the citation history of a cited reference.
Figure 3 shows a screenshot of the timeline visualization, in which clusters are displayed horizontally alone timelines. In timeline visualizations, the legend above the display area marks every 5 years. The label of each cluster is shown at the end of the cluster's timeline. Cited references or authors are depicted as circles filled with citation rings. The color of each ring corresponds to the time slice in which citations were made. The thickness of a ring is proportional to the amount of citations received in that time slice. Thus, a large-sized circle denotes a highly cited unit, i.e. reference or author. In timeline visualizations of cited authors, a cited author is positioned based on the earliest year in which he/she was cited in the dataset. A possible extension of this design would differentiate citations to the same author in different years.

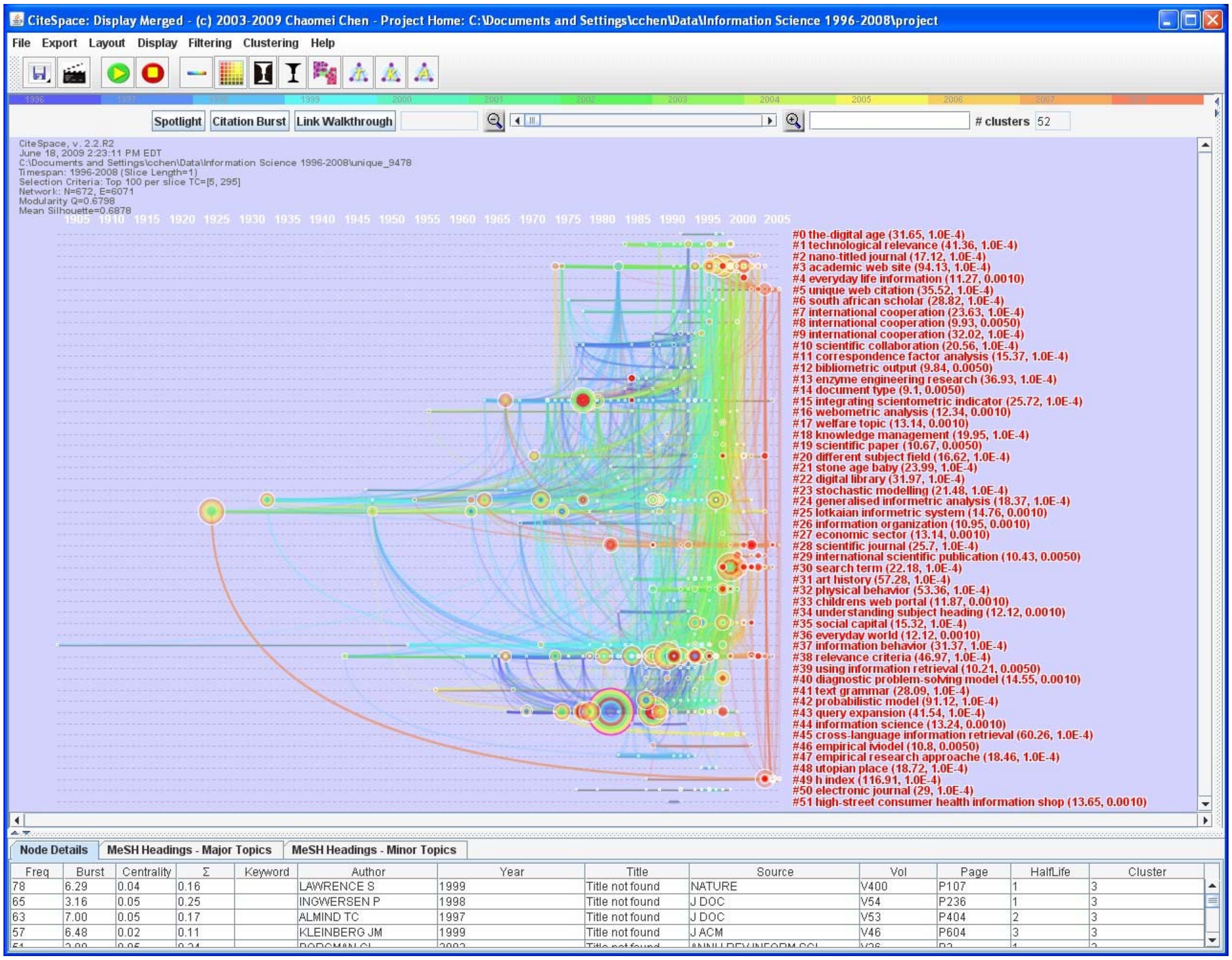

**Figure 3. The interactive visualization interface of *CiteSpace* showing a timeline visualization of clusters and their labels.**

Two additional colors, red and purple, are used to highlight special attributes of a node. A red ring indicates that a citation burst is detected in the corresponding time slice. A purple ring is added to a node if its betweenness centrality is greater than 0.1; the thickness of the ring is proportional to its centrality value.

A line connecting two items in the visualization represents a co-citation link. The thickness of a line is proportional to the strength of co-citation. The color of a line represents the time slice in which the co-citation was made for the first time. A useful byproduct of spectral clustering is that tightly coupled clusters tend to be placed next to each other and visually form a supercluster.

## Results

The results of the single-slice comparative ACA (2001-2005) are described first, followed by the findings of progressive ACA and DCA (1996-2008). The comparative ACA was limited to the 5-year period of 2001-2005 in order to synchronize with the time period of the study of Zhao and Strotmann (2008a). The full 13-year range of 1996-2008 was used in the other two progressive studies. Within each study, we summarize the prominent co-citation clusters in terms of their leading members, automatically generated labels based on information extracted from citing articles, and sentence summarization based on sentences in citing articles' abstracts.

*A Comparative ACA (2001-2005)*

The comparative ACA was conducted with reference to the ACA study of Zhao and Strotmann (2008a). Recall that Zhao and Strotmann identified 11 specialties based on 120 most cited authors in 2001-2005 and manually labeled these specialties by examining each specialty's members. They labeled one of factors as 'undefined.' We chose the top 120 most cited authors in the same time period using a single 5-year time slice in *CiteSpace*. Twelve author co-citation clusters were identified with a modularity of 0.5691, suggesting that inter-cluster connections are considerable but not overwhelming. The mean silhouette value of 0.7219 indicates a satisfactory partition of the network. The labels of these clusters were chosen from titles of their citers by *tf*idf* (See Figure 4). In contrast, Zhao and Strotmann (2008a) derived their labels from cited authors.

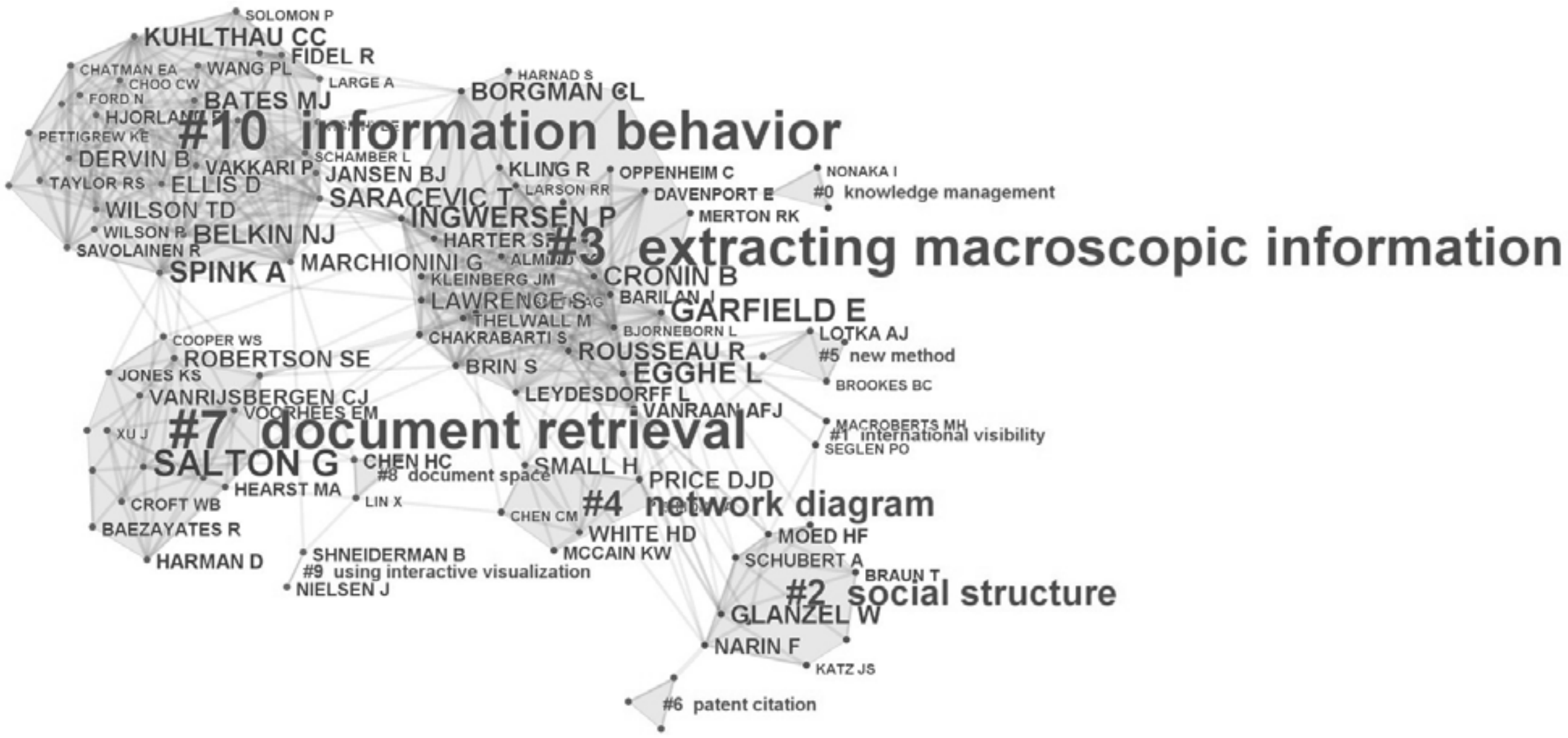

**Figure 4. A 120-author ACA network containing 11 clusters, generated with a single 5-year time slice (2001-2005). Clusters are labeled by citers' title terms using tf*idf weighting. An undefined cluster (#11) is omitted.**

Determining the number of specialties is a key issue in a co-citation analysis. In factor analysis, it is a common practice to identify specialties in terms eigenvectors with eigenvalues of 1 or greater. Zhao and Strotmann (2008a) also took into account other information such as the Scree plot, the total variance explained, communalities and correlation residuals. We tracked the number of clusters as the size of the ACA network increases (See Table 2). In general, the number of ACA clusters grows as the size of the ACA network increases. As shown in Figure 5, networks of 300~500 cited authors may be a good compromise with high values in both modularity and silhouette.

Table 2. The size of ACA networks and the number of ACA clusters. All use a single 5-year time slice.

| **# Cited Authors** | **% of Total Authors (24,648)** | **# Links** | **# ACA Clusters** | **Modularity Q** | **Mean Silhouette** |
|---|---|---|---|---|---|
| 60 | 0.24 | 234 | 4 | 0.4430 | 0.8202 |
| 110 | 0.45 | 469 | 7 | 0.5030 | 0.8293 |
| 120 | 0.49 | 514 | 12 | 0.5691 | 0.7219 |
| 200 | 0.81 | 915 | 20 | 0.6162 | 0.7974 |
| 300 | 1.22 | 1336 | 36 | 0.6864 | 0.6418 |
| 400 | 1.62 | 1712 | 46 | 0.7527 | 0.6763 |
| 500 | 2.03 | 2245 | 58 | 0.7718 | 0.6679 |
| 1000 | 4.06 | 5737 | 117 | 0.8725 | 0.5408 |

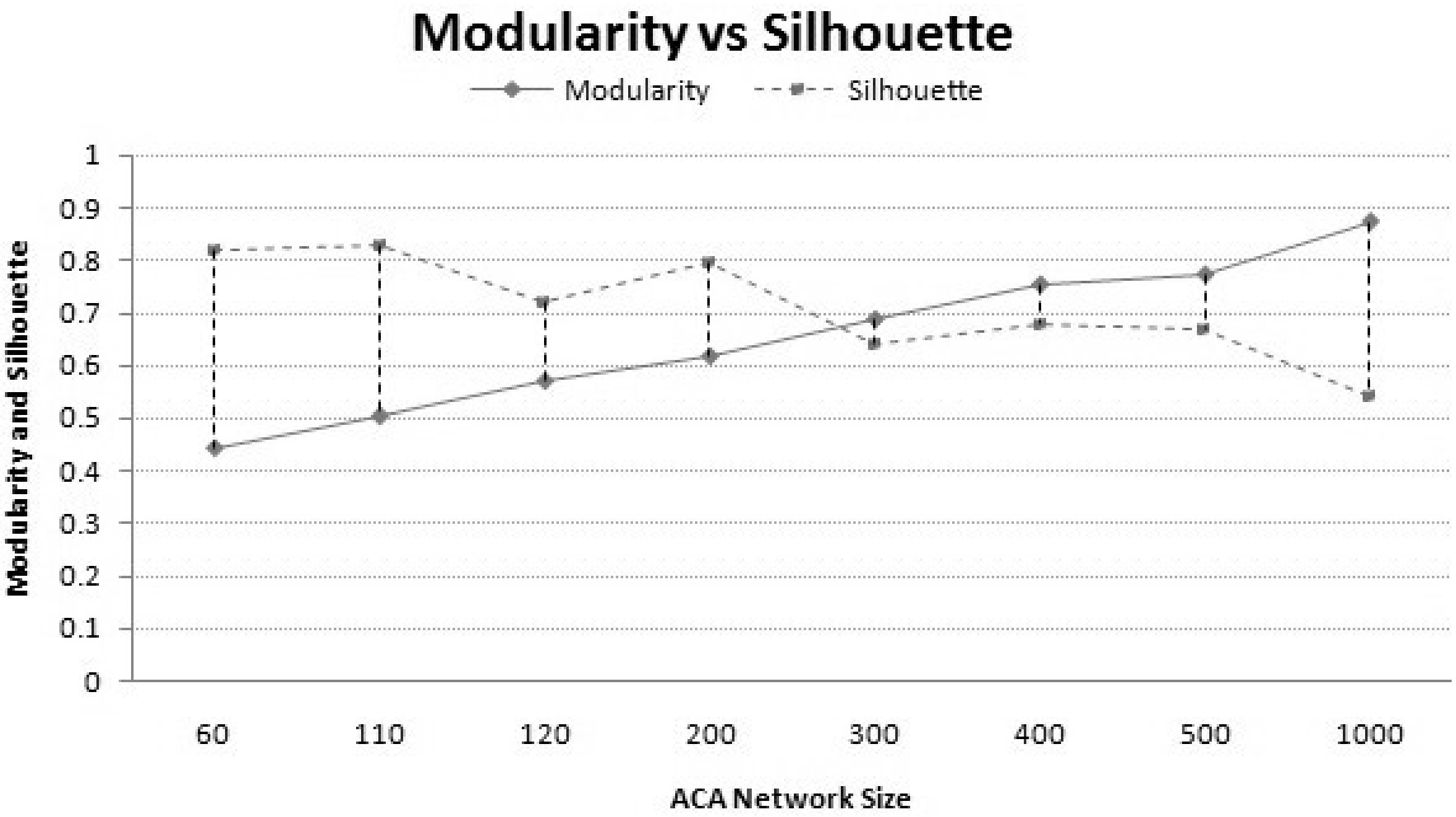

**Figure 5. The change of modularity and silhouette values as the network size increases.**

We compared 12 co-citation clusters ($C_i$) and 11 factors ($F_j$) in Zhao and Strotmann's ACA in terms of their overlapping members. Since each author can only appear in one cluster but may appear in multiple factors, one matching factor was selected only if the author has the greatest factor loading in absolute values; if no such factor was found, the author had no match. The overall overlapping rate is 82%, computed as follows:

$$\frac{\left|\left(\bigcup_{i=1}^{12} C_i\right) \cap \left(\bigcup_{j=1}^{11} F_j\right)\right|}{\left|\bigcup_{i=1}^{12} C_i\right|} = \frac{98}{120} = 0.82$$

Each cluster was projected as a distribution of its members over the 11 factors and the no-match category. Cluster $C_i$'s projection on factor $F_j$ is computed as: $\frac{|Ci \cap Fj|}{|C_i|}$. As shown in Table 3, cluster $C_3$'s projection on $F_{webometrics}$ is $\frac{|C_3 \cap F_{webometrics}|}{|C_3|} = \frac{19}{29} = 0.6552$.

Figure 6 depicts the overall matching patterns between clusters (diamonds) and factors (circles) in a similarity graph. The thickness (and darkness) of a line indicates the strength of the match. Figure 6 shows three types of patterns.

Type 1a-1c: a cluster primarily corresponds to a single factor, denoted as $C_i \Leftrightarrow F_j$.

(1a) $C_0 \Leftrightarrow F_{\text{knowledge management}}$

(1b) $C_3 \Leftrightarrow F_{\text{webometrics}}$

(1c) $C_7 \Leftrightarrow F_{\text{IR systems}}$

Type 2a-2b: two or more clusters are subsets of the same factor, i.e. for $K$ clusters, $\bigcup_{k=1}^{K} C_{i_k} \subseteq F_j$.

(2a) $C_1 \cup C_2 \cup C_4 \subseteq F_{scientometrics}$

(2b) $C_4 \cup C_8 \subseteq F_{mapping\ of\ science}$

Type 3a: one cluster is split into $L$ factors, i.e. $\bigcup_{l=1}^{L} F_{j_l} \subseteq C_i$.

(3a) $F_{information\ behavior} \cup F_{users\ judgements\ of\ relevance} \cup F_{childrens\ information\ behavior} \subseteq C_{10}$

Table 3. Authors in Cluster 3 are projected to the 11 factors and a no-match category.

| Cluster # | Zhao-Strotmann Factors | Matched Authors | Distribution (%) |
|---|---|---|---|
| 3 | webometrics | 19 | 65.52 |
| 3 | [no match] | 5 | 17.24 |
| 3 | scientometrics | 3 | 10.34 |
| 3 | children's info behavior | 1 | 3.45 |
| 3 | undefined | 1 | 3.45 |
| | Total | 29 | 100 |

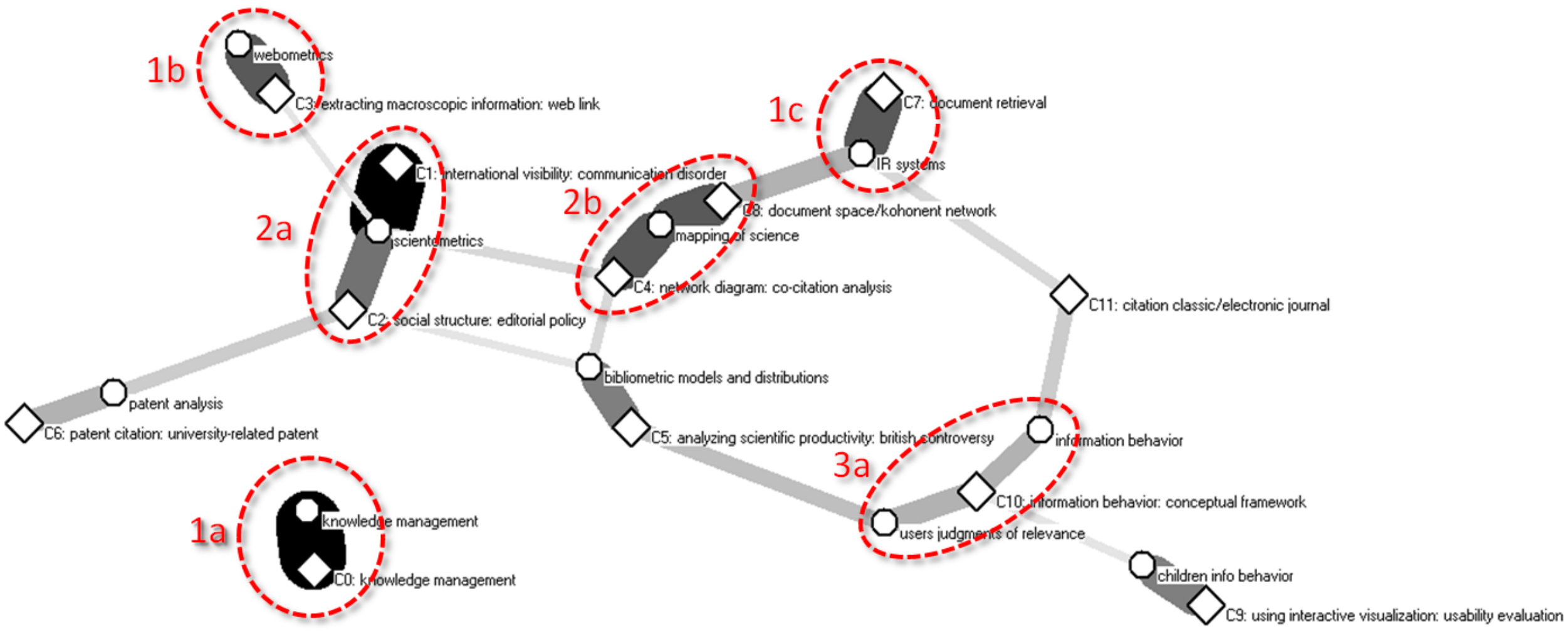

**Figure 6. An associative network of clusters (diamonds) and factors (circles) with 10% or more overlaps (thickness of line). Cluster labels are shown in two parts: terms chosen by tf*idf and by log-likelihood ratio.**

Factor labels given by Zhao and Strotmann tend to be conceptually higher-level terms than automatically generated cluster labels. For example, *patent analysis* is a broader term of *patent citation*; and *IR systems* is a broader term of *document retrieval*. Structurally, spectral clusters tend to be more specific groupings than factors. For instance, as shown in 2b, the *mapping of science* factor contains clusters such as $C_4$*: network diagram* (by tf*idf): *co-citation analysis* (by LLR), and $C_8$*: document space/Kohonen network*.
In summary, 1) spectral clustering and factor analysis identified about the same number of specialties, but they appeared to reveal different aspects of co-citation structures and 2) cluster labels chosen from citers of a cluster tend to be more specific terms than those chosen by human experts. These findings suggest that the multiple perspective method has the potential to provide additional insights in complementary to existing methods and provide an intermediate level of support for interpreting the nature of specialties.

*A Progressive ACA (1996-2008)*
This progressive ACA was a 13-year multiple-slice analysis of all the 5,963 records in types of article and review. A progressive co-citation analysis takes multiple co-citation networks from consecutive time intervals as input and produces a merged network to represent the evolution of the underlying domain (Chen, 2004). The inclusion of review-type records was to cover ARIST publications, which are classified as reviews. By including top-150 most cited authors from every year between 1996 and 2008, we obtained a merged network of 633 cited authors with 7,162 author co-citation links and 40 co-citation clusters. This 633-author network has a lower modularity (0.2278) than the smaller 120-author network in the comparative ACA (0.5691). Furthermore, the mean silhouette value (0.6929) of the larger network is also lower than that of the 120-author network (0.7219). The larger network has a much higher inter-cluster connectivity.
Figure 7 shows a timeline visualization of the 40 ACA clusters with automatically generated cluster labels. The display shows labels of highly cited authors in major clusters only. These clusters appear to be more interpretable than clusters in the comparative ACA. Visually, one may identify a few superclusters at the granularity of Zhao and Strotmann's factors. For example, clusters $C_2$, $C_4$, $C_7$, and $C_8$ form a supercluster that corresponds to the *scientometrics* factor.

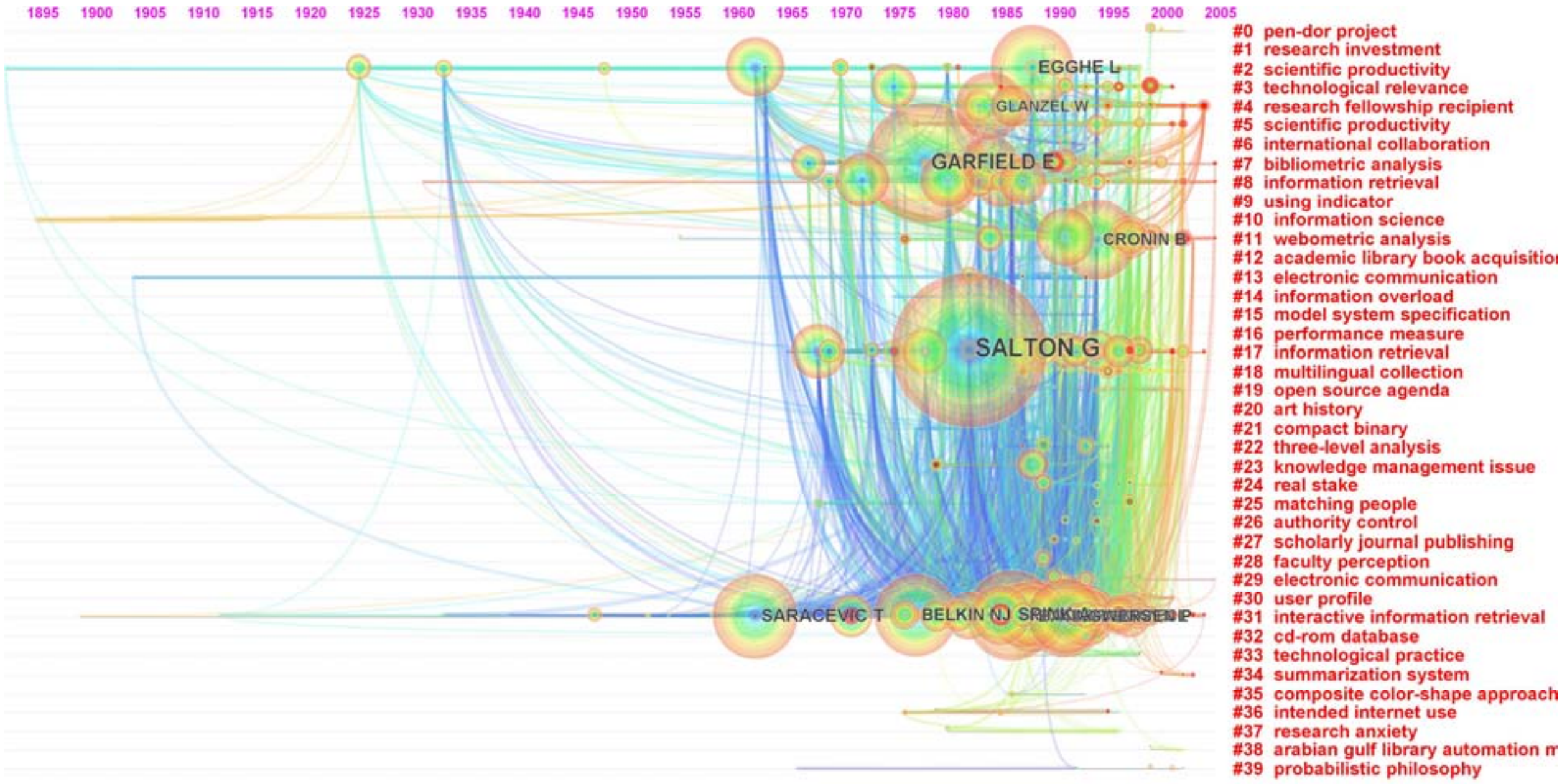

**Figure 7. 40 ACA clusters (1996-2008) (*CiteSpace* parameters: Nodes=633, Edges=7,162, top N=150, time slice length=1, modularity=0.2278, mean silhouette=0.6929, clusters=40).**

Table 4 shows automatically chosen cluster labels of the 6 largest ACA clusters along with their size and silhouette value. Top-ranked title terms by LLR were selected as cluster labels. The largest cluster *interactive information retrieval* (#31) has 199 members. Its negative silhouette value of -0.090 suggests a heterogeneous citer set. The second largest cluster (#17), with 95 members, is labeled as *information retrieval*. Other candidate labels for the cluster include *probabilistic model* and *query expansion*, confirming that this cluster deals with classic information retrieval issues. The third largest cluster (#7) is *bibliometric analysis*.

**Table 4. The 6 largest ACA clusters of a 633-author network (1996-2008).**

| # | N | Silhouette | Title terms by tf*idf | Title Terms by LLR (p=0.0001) |
|---|---|---|---|---|
| 31 | 199 | -0.090 | (80.30) interactive information retrieval<br>(62.46) information retrieval<br>(55.45) information science | (79.31) **interactive information retrieval**<br>(53.64) user information problem<br>(53.64) various aspect |
| 17 | 95 | 0.143 | (71.02) information retrieval<br>(41.44) probabilistic model<br>(37.94) query expansion | (68.19) **information retrieval**<br>(49.81) probabilistic model<br>(38.49) magazine article |
| 7 | 37 | 0.272 | (22.87) bibliometric analysis<br>(17.07) social science<br>(13.82) publication productivity | (31.51) **bibliometric analysis**<br>(22.79) social science<br>(22.67) career path |
| 2 | 33 | 0.343 | (20.97) scientific productivity<br>(14.98) statistical analysis<br>(14.76) analyzing scientific productivity | (24.24) theoretical population genetic<br>(24.16) **statistical analysis**<br>(23.54) new method |
| 11 | 32 | 0.682 | (20.72) webometric analysis<br>(18.44) informetric purpose<br>(18.44) data collection method | (32.65) **webometric analysis**<br>(28.16) data collection method<br>(28.16) informetric purpose |
| 8 | 30 | 0.330 | (19.68) information retrieval<br>(18.42) **citation analysis**<br>(16.12) information retrieval area | (34.42) **journal co-citation analysis**<br>(34.42) intellectual space<br>(34.42) information retrieval area |

Table 5 lists top-10 most cited authors of the six largest ACA clusters, including Spink_A and Saracevic_T in *interactive information retrieval* (#31), Salton_G, Robertson_SE, and van Rijsbergen_CJ in *information retrieval* (#17), Garfield_E, Moed_HF, and Merton_RK in *bibliometric analysis* (#7), Egghe_L, Price_DJD, and Lotka_AJ in *statistical analysis* (#2). The *webometric analysis* cluster (#11) includes Cronin_B, Rousseau_R, and Lawrence_S. The *journal co-citation analysis* (#8) includes Small_H, Leydesdorff_L, and White_HD.

Note that one may reach different insights into the nature of a co-citation cluster if different sources of information are used. The cited members of a cluster define its intellectual base, whereas citers to the cluster form a research front. The major advantage of our approach is that it enables analysts to consider multiple aspects of the citation relationship from multiple perspectives.

**Table 5. The six largest ACA clusters with most cited members and major citers of each cluster.**

| Cluster # and Label by LLR | Citations | Burst | Centrality | Sigma | Author | Major Citers (some subtitles are omitted due to the limited space) |
|---|---|---|---|---|---|---|
| 31. interactive information retrieval | 329 | | 0.03 | | SPINK A | (19) Sutcliffe AG (2000) empirical studies of end-user information searching |
| | 311 | | 0.02 | | SARACEVIC T | |
| | 299 | | 0.04 | | INGWERSEN P | |
| | 299 | | 0.02 | | BELKIN NJ | |
| | 272 | | 0.06 | | BORGMAN CL | (17) Robins D (2000) shifts of focus on various aspects of user information problems during interactive information retrieval<br>(17) Song M (2000) visualization in information retrieval |
| | 269 | | 0.04 | | BATES MJ | |
| | 239 | | 0.06 | | KUHLTHAU CC | |
| | 228 | 6.12 | 0.03 | 0.15 | DERVIN B | |
| | 225 | | 0.02 | | MARCHIONINI G | |
| | 223 | | 0.03 | | ELLIS D | |
| 17. information retrieval | 551 | 5.63 | 0.05 | 0.19 | SALTON G | (19) Dominich S (2000) a unified mathematical definition of classical information retrieval<br>(16) Sparck-Jones K (2000) a probabilistic model of information retrieval: development and comparative experiments part 2<br>(11) Efthimiadis EN (2000) interactive query expansion: a user-based evaluation in a relevance feedback environment |
| | 205 | | 0.02 | | ROBERTSON SE | |
| | 174 | 3.51 | 0.02 | 0.16 | VANRIJSBERGEN CJ | |
| | 166 | | 0.03 | | HARMAN D | |
| | 140 | 2.86 | 0.03 | 0.19 | JONES KS | |
| | 130 | | 0.01 | | VOORHEES EM | |
| | 128 | 4.62 | 0.03 | 0.16 | CROFT WB | |
| | 127 | | 0.01 | | HEARST MA | |
| | 117 | | 0.02 | | CHEN HC | |
| | 111 | 5.42 | 0 | 0.05 | BAEZAYATES R | |
| 7. bibliometric analysis | 429 | | 0.01 | | GARFIELD E | (7) Andersen H (2000) influence and reputation in the social sciences<br>(6) Case DO (2000) how can we investigate citation behavior?<br>(6) Dietz JS (2000) using the curriculum vita to study the career paths of scientists and engineers<br>(6) Prpic K (2000) the publication productivity of young scientists |
| | 213 | | 0.01 | | MOED HF | |
| | 139 | | 0.02 | | MERTON RK | |
| | 103 | | 0.01 | | SEGLEN PO | |
| | 101 | 9.19 | 0.01 | 0.07 | CASE DO | |
| | 89 | 3.87 | 0.01 | 0.12 | MACROBERTS MH | |
| | 86 | | 0.04 | | LATOUR B | |
| | 77 | | 0.01 | | VINKLER P | |
| | 75 | | 0.01 | | COLE S | |
| | 68 | 3.03 | 0.02 | 0.16 | BUDD JM | |
| 2. statistical analysis(2) | 302 | | 0.01 | | EGGHE L | (6) Egghe L (2000) aging, obsolescence, impact, growth, and utilization: definitions and relations<br>(6) Gupta BM (2000) modelling the growth of literature in the area of theoretical population genetics |
| | 215 | | 0.03 | | PRICE DJD | |
| | 103 | | 0.01 | | LOTKA AJ | |
| | 82 | | 0.02 | | BRADFORD SC | |
| | 73 | | 0.01 | | BURRELL QL | |
| | 72 | | 0.02 | | CRANE D | |
| | 72 | 4.15 | 0.01 | 0.10 | LINE MB | |
| | 71 | 3.06 | 0.02 | 0.17 | PAO ML | |
| | 62 | 3.36 | | 0.01 | ZIPF GK | |
| | 46 | 4.80 | | 0.06 | GRIFFITH BC | |
| 11. webometric analysis | 294 | | 0.03 | | CRONIN B | (8) Thomas O (2000) webometric analysis of departments of librarianship and information science<br>(6) Bar-Ilan J (2000) the web as an information source on informetrics? a content analysis |
| | 215 | | 0.01 | | ROUSSEAU R | |
| | 158 | 2.91 | 0.01 | 0.11 | LAWRENCE S | |
| | 129 | 3.81 | 0.01 | 0.10 | BRIN S | |
| | 113 | | 0.01 | | OPPENHEIM C | |
| | 110 | | 0.01 | | BARILAN J | |
| | 109 | 7.03 | 0.01 | 0.06 | THELWALL M | |
| | 78 | | 0.01 | | DAVENPORT E | |
| | 75 | | 0.01 | | CHEN CM | |
| | 72 | 8.52 | 0.01 | 0.05 | VAUGHAN L | |
| 8. journal co-citation analysis | 205 | | 0.02 | | SMALL H | (10) Ding Y (2000) bibliometric information retrieval system (birs): a web search interface utilizing bibliometric research results<br>(7) Ding Y (2000) journal as markers of intellectual space: journal co-citation analysis of information retrieval area, 1987-1997 |
| | 202 | | 0.05 | | LEYDESDORFF L | |
| | 202 | | 0.07 | | WHITE HD | |
| | 183 | | 0.02 | | VANRAAN AFJ | |
| | 130 | | 0.04 | | MCCAIN KW | |
| | 90 | | 0.02 | | CALLON M | |
| | 81 | | 0.01 | | KOSTOFF RN | |
| | 75 | 4.32 | 0.01 | 0.07 | KUHN TS | |
| | 69 | | | | PERSSON O | |
| | 55 | | | | NOYONS ECM | |

*The Progressive DCA (1996-2008)*

In the progressive DCA, co-citation networks were first constructed with the top-100 most cited documents in each of the 13 one-year time slices between 1996 and 2008. Then, these networks were merged into a network of 655 co-cited references. The merged network was subsequently decomposed into 50 clusters. Table 6 summarizes these clusters. We first provide an overview of these clusters and discuss the five largest clusters in detail.

**Overview**

The 50 clusters vary considerably in size. The largest cluster #18 contains 150 members, which is 22.90% of the entire set of 655 references. The five largest clusters altogether reach 51.60%. In contrast, there are six clusters contain only two members.

The network's overall mean silhouette value is 0.7372, which is the highest among the three co-citation networks we analyzed in this article. In general, the silhouette value of a cluster is negatively correlated with its size (-0.654). For example, the largest cluster, #18, has the lowest silhouette value of -0.024, indicating its diverse and heterogeneous structure. In contrast, the second largest cluster, #43, has a more homogenous structure with a reasonably high silhouette value of 0.522. The fifth largest cluster, #2, has a very high silhouette value of 0.834. The following discussion will focus on the five largest clusters and their interrelationships.

**Table 6. The five largest clusters sorted by size.**

| | C# | n | % | Silhouette | Title Terms (tf*idf) | Title Terms (LLR) (*p=0.0001) |
|---|---|---|---|---|---|---|
| 1 | 18 | 150 | 22.90 | -0.024 | (80.82) **interactive information retrieval**<br>(72.92) information retrieval<br>(51.5) user information problem | **interactive information retrieval** (294.13*)<br>user information problem (167.06*)<br>various aspect (167.06*) |
| 2 | 43 | 69 | 10.53 | 0.522 | (131.97) **academic web**<br>(103.62) web site<br>(54.77) exploratory hyperlink | **academic web** (174.79*)<br>exploratory hyperlink (152.94*)<br>linguistic consideration (152.94*) |
| 3 | 13 | 46 | 7.02 | 0.153 | (42.16) **information retrieval**<br>(23.47) probabilistic model<br>(22.53) query expansion | **information retrieval** (104.03*)<br>probabilistic model (81.54*)<br>using heterogeneous thesauri (67.95*) |
| 4 | 35 | 44 | 6.72 | 0.245 | (14.07) **citation behavior**<br>(14.07) citing literature<br>(11.74) citation theory | **citation behavior** (56.66*)<br>citing literature (56.66*)<br>citation theory (43.92*) |
| 5 | 2 | 29 | 4.43 | 0.834 | (83.69) **h index** | **h index** (212.76*) |
| | | | | | (53.18) successive h-indices<br>(43.03) generalized hirsch h-index | generalized hirsch h-index (156.63*)<br>disclosing latent fact (156.63*) |

**Five Major Clusters: Cited References as the Intellectual Base**

The five largest document co-citation clusters are *interactive information retrieval* (#18), *academic web* (#43), *information retrieval* (#46), *citation behavior* (#44), and *h index* (#2). We analyzed two aspects of each specialty: 1) prominent members of a cluster as the intellectual basis and 2) themes identified in the citers of the cluster as research fronts.

Table 7 summarizes top-cited cluster members and their structural, temporal, and saliency metrics such as citation count (**φ**), betweenness centrality (**σ**), citation burstness (**τ**), and sigma – a novelty indicator (∑) (Chen et al., 2009a). The most cited references in the *interactive information retrieval* cluster are Kuhlthau_1991 on information seeking and Bates_1989 on browsing and berrypicking. The first four references are related to information needs and user-centric approaches. However, we did not expect to see the 5$^{th}$ reference, the ACA study by White_1998, because in our mind it firmly belongs to a citation analysis specialty. On the other hand, the low silhouette value of the *interactive information retrieval* cluster suggests that a diverse and complex range of research fronts may have drawn upon the knowledge of this cluster. The following discussion on the second aspect – themes identified among citers – may give us additional insights into the nature of this cluster.

The stars in the *academic web* cluster (#43) are Lawrence_1999 and Kleinber_1999. Both papers were published outside the domain defined by the 12 source journals; instead, they appeared in *Nature* and *JACM*. This is an example of how one discipline (information science) was influenced by another (computer science). It would explain why the cluster was dominated by the work of '*outsiders*.' In contrast, the *information retrieval* cluster (#13) is, as one would expect, dominated by the work of '*insiders*,' namely, books by Salton and by van Rijsbergen. Similarly, the *citation behavior* cluster (#35) was dominated by '*insiders*,' for example, Garfield's book on citation indexing, Small's 1973 paper on co-citation, Price's 1965 paper on networks of scientific papers, and White and Griffith's 1981 paper introducing author co-citation. A key that may distinguish *insiders* from *outsiders* is that insiders may cite the original cluster, but outsiders will probably not.

The core of the fifth largest cluster, the *h index* cluster (#2), is Hirsch_2005, which originally introduced the concept of h-index. The strongest citation burst of 15.75 was detected in the citation history of Hirsch_2005. As our analysis will demonstrate, the *h index* cluster is one of the most active areas of research in recent years.

The average age of core papers in a cluster is an estimation of the time the cluster was formed. According to the average age of top-5 core papers, the 37-year old *citation* cluster is the oldest – formed around 1973, its average year of publication, whereas the *h-index* cluster is the youngest – 5 years old, formed in 2005. In between, the *information retrieval* cluster (#13) is 31 (formed in 1979); the *interactive information retrieval* cluster (#18) is 18 formed in 1992; and the *academic web* cluster (#43) is 11 (formed in 1999).

Table 7. Most frequently cited references in the five largest document co-citation clusters ranked by citation counts φ.

| Cluster # | φ | τ | σ | ∑ | Cited References |
|---|---|---|---|---|---|
| 18 | 118 | 0.00 | 0.04 | 0.00 | Kuhlthau, C. C. (1991) Inside the search process: Information seeking from the user's perspective. *Journal of the American Society for Information Science*, 42 (5), 361-371. |
| | 95 | 4.79 | 0.04 | 0.19 | BATES MJ (1989) The design of browsing and berrypicking techniques for the online search interface. ONLINE REV, 13, 407. |
| | 85 | 8.52 | 0.03 | 0.11 | Dervin, B. and Nilan, M. (1986). Information needs and uses. *Annual Review of Information Science and Technology* (ARIST), v. 21, pp 3-33. |
| | 81 | 0.00 | 0.03 | 0.00 | INGWERSEN P (1996) Cognitive perspectives of information retrieval interaction: elements of a cognitive IR theory. J DOC, 52(1), 3-50. |
| | 79 | 4.04 | 0.04 | 0.21 | WHITE HD and McCain KW (1998) Visualizing a discipline: An author co-citation analysis of information science, 1972-1995. J AM SOC INFORM SCI, 49, 327. |
| 43 | 76 | 8.83 | 0.06 | 0.17 | LAWRENCE S (1999) Accessibility and distribution of information on the Web, Nature, 400, 107. |
| | 64 | 9.00 | 0.06 | 0.16 | ALMIND TC (1997) Informetric analyses on the world wide web: methodological approaches to 'Webometrics', J DOC, 53, 404. |
| | 63 | 3.44 | 0.04 | 0.22 | INGWERSEN P (1998) The calculation of Web impact factors. J DOC, 54, 236 |
| | 53 | 6.58 | 0.02 | 0.12 | Kleinberg, J. M. (1999) Authoritaive sources in a hyperlinked environment. JACM, 46, 604-632. |
| | 50 | 7.12 | 0.03 | 0.13 | Rob Kling and Geoffrey W. McKim (2000) Not just a matter of time: Field differences and the shaping of electronic media in supporting scientific communication. JASIS, 51(14), 1306-1320. |
| 13 | 153 | 11.38 | 0.05 | 0.13 | SALTON G (1983) Introduction to modern information retrieval. |
| | 80 | 3.21 | 0.07 | 0.29 | VANRIJSBERGEN CJ (1979) Information Retrieval |
| | 75 | 0.00 | 0.03 | 0.00 | ROBERTSON SE (1976) Relevance weighting of search terms, J AM SOC INFORM SCI, 27, 129 |
| | 72 | 0.00 | 0.02 | 0.00 | Salton, G., and Buckley, C. (1988) Term-Weighting Approaches in Automatic Text Retrieval, Information Processing & Management, 24(5), 513-523. |
| | 44 | 0.00 | 0.01 | 0.00 | ROCCHIO JJ (1971) Relevance Feedback in Information Retrieval, chapter 14 SMART retrieval system, pp. 313-323 |
| 35 | 86 | 12.26 | 0.09 | 0.17 | GARFIELD E (1979) Citation indexing: Its theory and application in science, technology, and humanities |
| | 75 | 0.00 | 0.05 | 0.00 | Small, H. (1973) Cocitation in scientific literature: New measure of relationship between 2 documents. Journal of the American Society for Information Science, 24(4), 265-269. |
| | 62 | 0.00 | 0.08 | 0.00 | Derek J. de Solla Price (1965) Networks of scientific papers: The pattern of bibliographic references indicates the nature of the scientific research front, Science, 149(3683), 510-515. |
| | 61 | 0.00 | 0.06 | 0.00 | White, H.D. & Griffith, B.C. (1981). Author cocitation: A literature measure of intellectual structure. Journal of the American Society for Information Science, 32, 163-171. |
| | 61 | 0.00 | 0.03 | 0.00 | MERTON RK (1968) The Matthew effect in science: The reward and communication systems of science are considered, Science, 159, 56. |
| 2 | 42 | 15.75 | 0 | 0.02 | HIRSCH JE (2005) An index to quantify an individual's scientific research output, P NATL ACAD SCI USA, 102, 16569 |
| | 24 | 8.98 | 0 | 0.01 | Bornmann, L. & Daniel, H.-D. (2005) Does the *h*-index for ranking of scientists really work? Scientometrics, 65(3), 391-392. |
| | 22 | 7.54 | 0 | 0.02 | Ball, P. (2005) Index aims for fair ranking of scientists, NATURE, 436(7053), 900. |
| | 19 | 7.11 | 0 | 0.01 | Branu, Tibor (2005) A Hirsch-type index for journals, The Scientists, 19(22), 8. |
| | 18 | 6.73 | 0 | 0.02 | Egghe, L. (2005). Power laws in the information production process: Lotkaian informetrics. Elsevier: Oxford, UK. |

**Five Major Clusters: Citing Articles as Research Fronts**

Research fronts of a document co-citation cluster were characterized by terms extracted from the citers of the cluster. Nine methods of ranking extracted terms were implemented in *CiteSpace* by choosing terms from three sources – titles, abstracts, and index terms of the citers of each cluster – and three ranking algorithms, namely, tf*idf weighting (Salton et al., 1975), log-likelihood ratio tests (LLR) (Dunning, 1993; Witten & Frank, 1999), and mutual information (MI)(Witten & Frank, 1999). Top-ranked terms became candidate cluster labels.

The reliability of these term ranking methods was measured by a consensus score $r = 0.1*(n + 1)$, where n is the number of other methods that also top rank the same term. It turned out that the best three ranking methods were: 1) title terms ranked by LLR, 2) index terms ranked by LLR, and 3) title terms ranked by tf*idf (See Figure 8). tf*idf and LLR produced identical labels for 36 clusters out of 50 (72%). Table 8 summarizes candidate labels chosen by different methods.

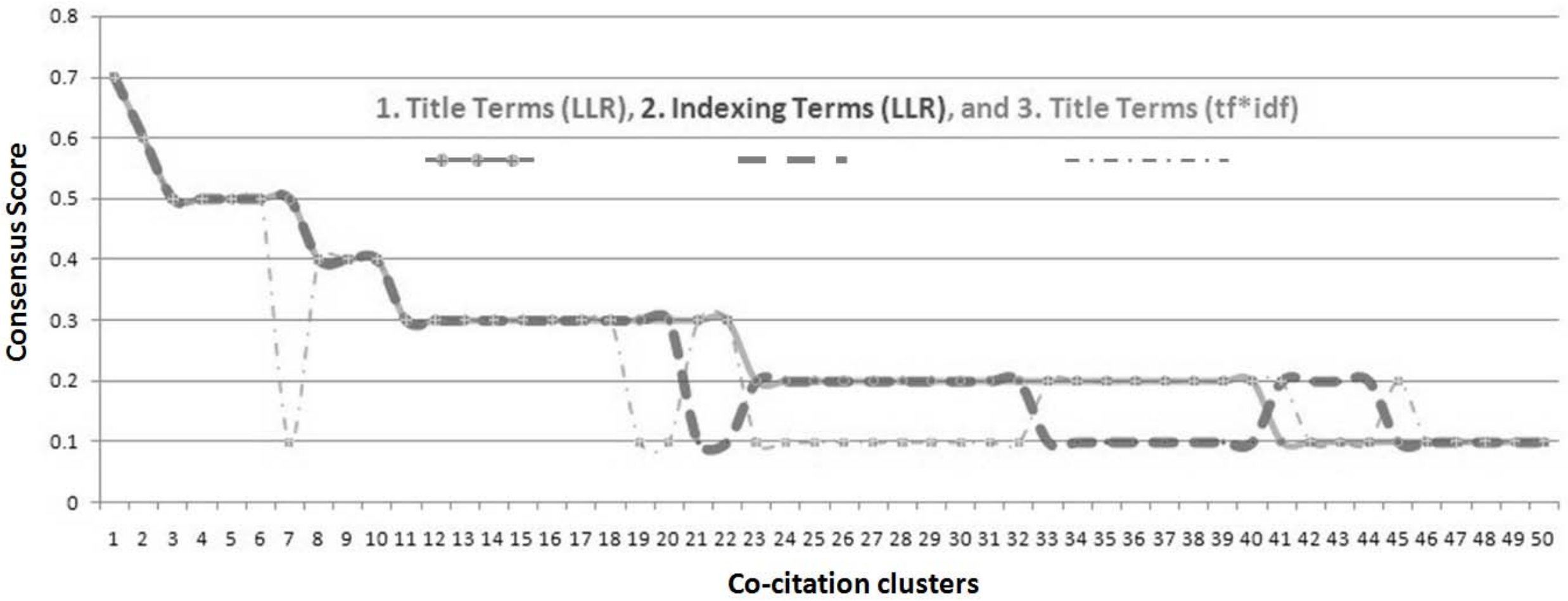

**Figure 8. The best consensus-conforming ranking methods: 1) title terms by LLR, 2) index terms by LLR, and 3) title terms by tf*idf.**

We discussed earlier that the largest cluster (#18) has 150 members and it has the lowest silhouette value (See Table 6). It turned out that the cluster was cited by 185 citing articles in the dataset. A total of 869 terms were extracted from the titles of these citing articles. In order to verify the heterogeneity of this set of citers, the term similarity network was decomposed using singular value decomposition (SVD). As a result, the term space was indeed multi-dimensional in nature because the largest connected component of the term similarity network contains only 353 terms, which is 40.62% of the 869 terms. In contrast, the *h index* cluster (#2) was much more homogeneous; the cluster was the citation footprint of 39 citing articles. Table 8 includes the silhouette value of each cluster.

**Table 8. The candidate cluster label terms selected by 6 of the 9 methods. The numbers under term sources are the consensus scores of the corresponding methods. The 3 mutual information methods are not shown due to lower scores (6.8, 6.8, 6.1 for index terms, titles, and abstracts). The most popular terms across the six methods are chosen to be the cluster labels.**

| | | Top 3 Terms by tf*idf | | | Top 3 Terms by LLR *p=0.0001 | | |
|---|---|---|---|---|---|---|---|
| # | Silhouette | Index (5.5) | Title (11.6) | Abstract (8) | Index (12.4) | Title (13.2) | Abstract (9.8) |
| 18 | -0.024 | (124.75) design (124.01) relevance (105.72) systems | (80.82) **interactive information retrieval** (72.92) information retrieval (51.5) user | (281.15) ir system (225.65) query expansion (176.8) information -seeking | **interactive information retrieval** (435.71*) design (336.26*) relevance (316.45*) | **interactive information retrieval** (294.13*) user information problem (167.06*) | **interactive information retrieval** (583.69*) information -seeking strategy |
| | | | information problem | strategy; | | various aspect (167.06*) | (510.18*) relevance criteria (436.46*) |
| 43 | 0.522 | (138.92) webometrics (133.86) links (129.76) bibliometrics | (131.97) **academic web** (103.62) web site (54.77) exploratory hyperlink | (210.46) web site (163.18) university web (154.74) research productivity | impact factors (394.55*) bibliometrics (299.87*) webometrics (274.41*) | **academic web** (174.79*) exploratory hyperlink (152.94*) linguistic consideration (152.94*) | web page (482.53*) impact factors (451.5*) research productivity (399.45*) |
| 13 | 0.153 | (41.29) document-retrieval (35.36) inference (33.96) retrieval | (42.16) **information retrieval** (23.47) probabilistic model (22.53) query expansion | (78.64) end user (61.16) expansion term (57.56) query expansion | **information retrieval** (166.64*) probabilistic model (127.52*) using heterogeneous thesauri (106.35*) | **information retrieval** (104.03*) probabilistic model (81.54*) using heterogeneous thesauri (67.95*) | **information retrieval** (220.04*) end user (216.55*) probabilistic model (199.15*) |
| 35 | 0.245 | (15.73) scientific literatures (15.48) journals (14.66) scientific literature | (14.07) **citation behavior** (14.07) citing literature (11.74) citation theory | (49.15) end user (31.8) search domain (28.97) citation theory | **citation behavior** (96.24*) citing literature (96.24*) indicator theory (72.87*) | **citation behavior** (56.66*) citing literature (56.66*) citation theory (43.92*) | citation analysis (119*) citation theory (115.11*) end user (111.69*) |
| 2 | 0.834 | (84.4) hirsch-index (82.31) ranking (80.47) scientists | (83.69) **h index** (53.18) successive h-indice (43.03) generalized hirsch h-index | (280.04) **h index** (53.04) scientific research (48.28) new index | **h index** (317.99*) ranking (317.24*) scientists (301.41*) | **h index** (212.76*) generalized hirsch h-index (156.63*) disclosing latent fact (156.63*) | **h index** (852.06*) ranking (310.43*) scientists (294.91*) |

The second largest cluster was labeled as *academic web* by LLR, but the top-ranked index term was *webometrics*, which was also the name of a specialty identified by Zhao and Strotmann. The index term *webometrics* is broader and more generic than the term *academic web*. This observation suggests that a manual labeling process is probably very similar to the indexing process after all.

The identification of the *h-index* cluster is unique because there was no such cluster in the 1996-2008 ACA. This is a good example why one should consider both ACA and DCA so that distinct DCA clusters such as the *h-index* one can be detected.

The time span $\tau$ between a research front and its intellectual base can be estimated as the difference between their average years of publications:

$$\tau(C_i) = \frac{\sum_{d \epsilon citers(C_i)} year(d)}{|citers(C_i)|} - \frac{\sum_{c \epsilon C_i} year(c)}{|C_i|} + 1$$

For example, *citation behavior* (#35) has the longest time span, $\tau(C_{35})$ = 2000-1973=28 years. IR has the second longest time span $\tau(C_{13})$ = 2000-1979=22. The time span for *interactive IR* is $\tau(C_{18})$ = 2000-1992=9 years; for *academic web* (#43), $\tau(C_{43})$ = 2003-1999=5 years; and for *h-index* (#2), $\tau(C_2)$ = 2007-2005=3 years.

Table 9 lists the most representative citing articles in each cluster. For example, Thelwall has a prominent role in the research front of the *academic web* cluster (#43). He authored 3 of the top 5 citing articles of the cluster, including Thelwall_2003 which cited 14 references of the cluster.

**Table 9. Titles of the two most frequent citers to each of the 5 largest DCA clusters. Terms chosen by LLR are underlined.**

| # | Cluster Label | Titles of Key Citers |
|---|---|---|
| 18 | Interactive information retrieval | (16) Robins D (2000) shifts of focus on various aspects of user information problems during <u>interactive information retrieval</u><br>(15) Beaulieu M (2000) <u>interaction</u> in information searching and <u>retrieval</u> |
| 43 | Academic web | (14) Thelwall M (2003) disciplinary and linguistic considerations for <u>academic web</u> linking: an exploratory hyperlink mediated study with mainland china and taiwan<br>(12) Wilkinson D (2003) motivations for <u>academic web</u> site interlinking: evidence for the web as a novel source of information on informal scholarly communication |
| 13 | Information retrieval | (8) Ding Y (2000) bibliometric <u>information retrieval</u> system (birs): a web search interface utilizing bibliometric research results<br>(6) Dominich S (2000) a unified mathematical definition of classical <u>information retrieval</u><br>(6) Sparck-Jones K (2000) a probabilistic model of <u>information retrieval</u>: development and comparative experiments part 2 |
| 35 | Citation behavior | (5) Case DO (2000) how can we investigate <u>citation behavior</u>? a study of reasons for citing literature in communication<br>(5) Ding Y (2000) bibliometric information retrieval system (birs): a web search interface utilizing bibliometric research results |
| 2 | H index | (14) Bornmann L (2007) what do we know about the <u>h index</u>?<br>(11) Sidiropoulos A (2007) generalized hirsch h-index for disclosing latent facts in citation networks |

**Automatic DCA Cluster Summarization**

Summarization sentences were ranked by the earlier described energy function *E* (Fernandez et al., 2007). Table 10 highlights two top-ranked summarization sentences for the same five DCA clusters we discussed earlier. Analysts can enhance their understanding of a cluster based on the additional contextual information provided by summarization sentences. The TreeTagger (Schmid, 1999) trained on the Penn Tree Bank, was used here for sentence segmentation.

### Table 10. The top two summarization sentences selected from the 5 largest DCA clusters.

| # | Highly ranked sentences |
|---|---|
| 18 | UT: 000166428400004 (E=2970848)<br><br>… (2) to show how the findings of the study refine Kuhlthau 's model of the **information search process in the field of information retrieval** (IR)… |
| | UT: 000206501000004 (E=2634203)<br><br>Design/ methodology/ approach - The paper outlines and explores a geographic **information seeking** process in which geographic information needs ( conditioned by needs and tasks, in context) drive the acquisition and use of geographic information objects, which in turn influence geographic behaviour in the environment. |
| 43 | UT: 000169989400001 (E=242241)<br><br>In this paper, we analyze two views of information production and use in computer-related research based on citation analysis of PDF and Postcript formatted **publications on the Web** using autonomous citation indexing (ACI), and a parallel citation analysis of the journal literature indexed by the Institute for Scientific Information (ISI) in SCISEARCH. |
| | UT: 000169392900004 (E=237239)<br><br>Google 's web ranking algorithm (Brin & Page, 1998) on **ranking web pages** is applied in this new coordinate space. |
| 13 | UT: 000088865500003 (E=420692)<br><br>(3) The relationships identified between the five best terms selected by the users for **query expansion** and the initial **query terms** were that: (a) 34 % of the **query expansion terms** have no relationship or other type of correspondence with a query term; (b) 66 % of the remaining **query expansion terms** have a relationship to the query terms. |
| | UT: 000085411400002 (E=349431)<br><br>We provide data on: (i) sessions - changes in queries during a session, number of pages viewed, and use of relevance feedback; (ii) queries - the number of **search terms**, and the use of logic and modifiers; and (iii) terms - their rank/ frequency distribution and the most highly used **search terms**. |
| 35 | UT: 000088918500005 (E=281345)<br><br>A **journal co-citation analysis** of fifty journals and other publications in the information retrieval (IR) discipline was conducted over three periods spanning the years of 1987 to1997. |
| | UT: 000246678500011 (E=269105)<br><br>We use a new data gathering method, "Web/URL citation," Web/ URL and Google Scholar to **compare traditional and Web-based citation patterns** across multiple disciplines ( biology, chemistry, physics, computing, sociology, economics, psychology, and education) based upon a sample of 1,650 articles from 108 open access ( OA) journals published in 2001. |
| 2 | UT: 000246678500011 (E=133747)<br><br>We use a new data gathering method, " Web/ URL citation, " Web/ URL and Google Scholar to **compare traditional and Web-based citation patterns** across multiple disciplines (biology, chemistry, physics, computing, sociology, economics, psychology, and education) based upon a sample of 1,650 articles from 108 open access (OA) journals published in 2001. |
| | UT: 000232418400010 (E=131078)<br><br>Hirsch ( 2005) has proposed the **h-index as a single-number criterion to evaluate the scientific output of a researcher** (Ball, 2005). |

The *h index* cluster (#2) is the most active one in terms of citation burstness. The top 10 sentences from this cluster represented themes such as studies of the h-index and its variations (the 2nd-4th, 7th-8th and 10th sentences), the studies of web-based citation patterns (the 1st sentence), and new resources of scientometrics, i.e. Scopus and Google Scholar (the 5th, 6th, and 9th sentences).
Considering that the system only has abstracts rather than full-texts to work with, the results are very encouraging because selected sentences will largely increase the ability of an analyst, or even a layman, to comprehend the meaning and significance of clusters in citation analysis much more easily.

**Visual Exploration**
The DCA network shown in Figure 9 was generated by *CiteSpace*. The 655 references and 6,099 co-citation links were divided into 50 clusters with a modularity of 0.6205, which represents a considerable amount of inter-cluster links. Major clusters are labeled in the visualization in red color with the font size proportional to the size of clusters. The colors of co-citation links reveal that the earliest inter-cluster connection is between *interactive IR* and *IR*, followed by connections between *interactive IR* and *citation behavior*, then by the more recent connections between *academic web*, *citation behavior*, and *IR*, and finally, the most recent connections between *h index*, *citation behavior*, *academic web*.

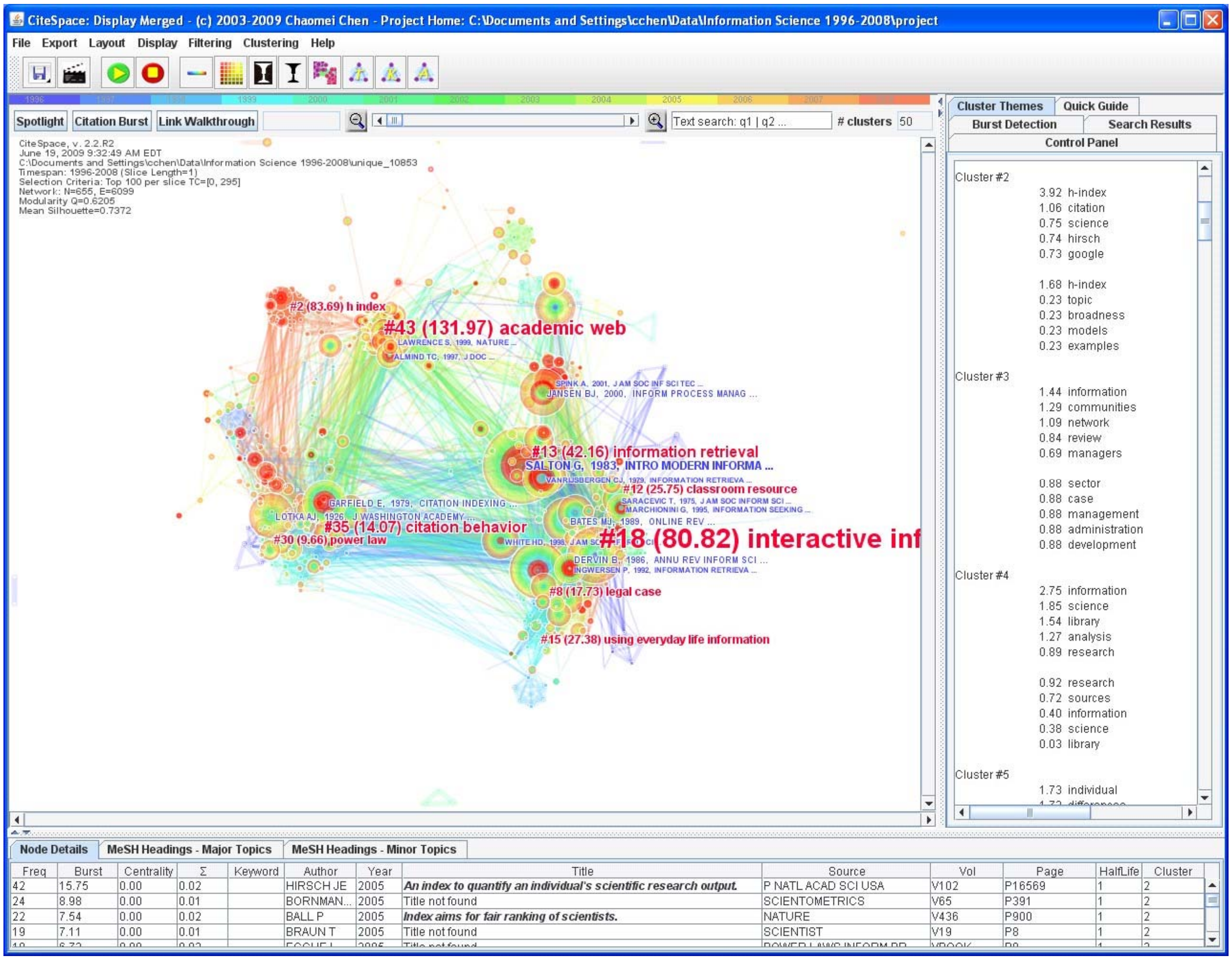

**Figure 9. An overview of the co-citation networks. Cited references with highest sigma values are labeled.**

Figure 10 shows a timeline visualization of the 50 clusters and their interrelationships. Each cluster is plotted horizontally. Each timeline runs from left to right with its label displayed to the right. The design of the timeline visualization is inspired by the work of Morris, Yen, Wu, and Asnake (2003) and further enhanced by automatic clustering labeling with multiple algorithms. Analysts can visually identify a variety of characteristics of a cluster, such as the length of its history, its citation classics, citation bursts, and how it is connected to other clusters. For example, in the citation behavior cluster, Garfield's 1979 book on citation index stands out with the highest betweenness centrality of 0.09. Many large citation circles would identify a high-impact specialty, whereas many red rings of citation bursts would highlight emerging specialties.

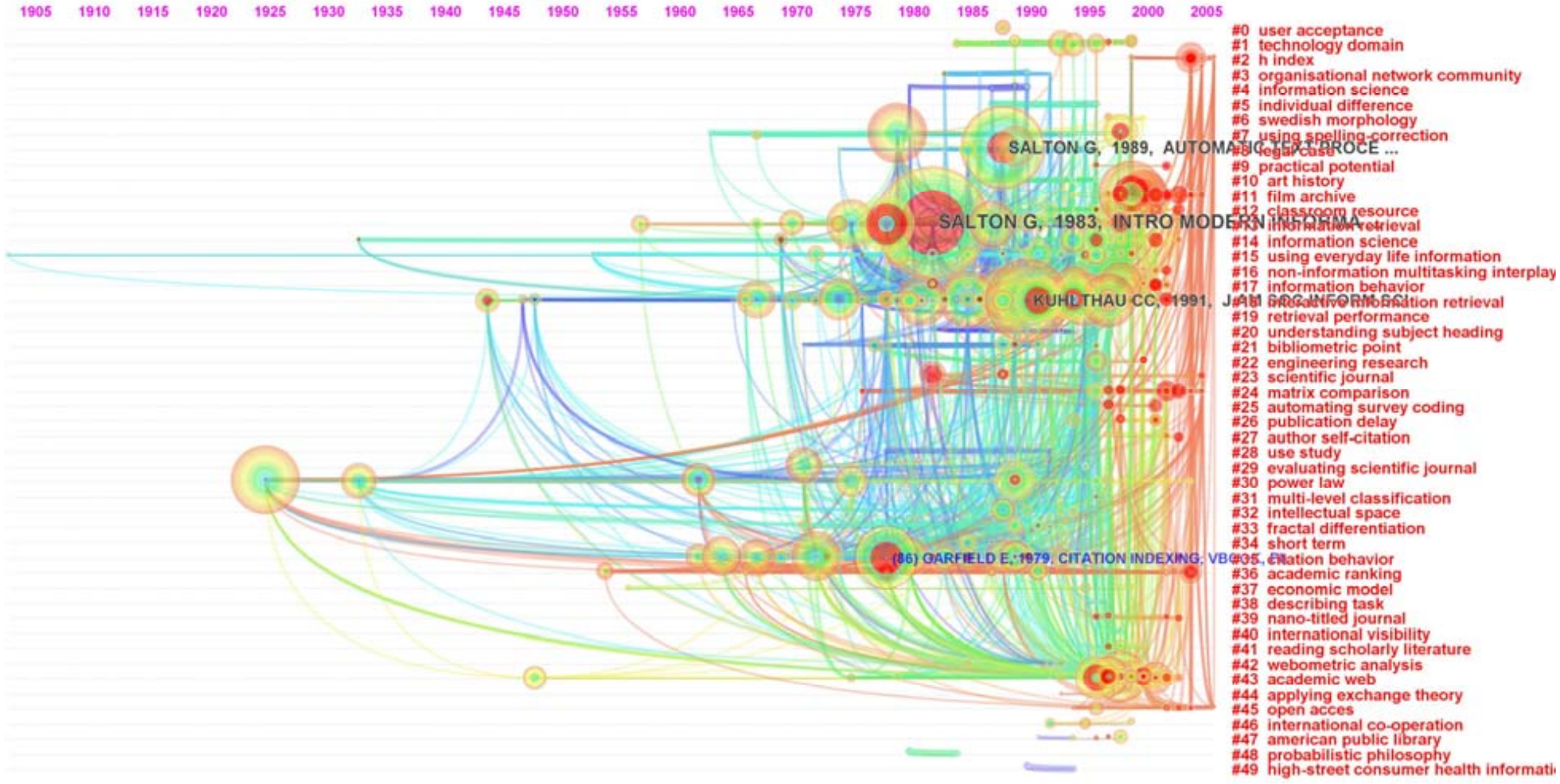

**Figure 10. A timeline visualization of the 50 DCA clusters (655 nodes, 6,099 links, modularity=0.6205, mean Silhouette=0.7372). Cluster labels are automatically generated from title terms of citing articles of specific clusters.**

It is clear in the timeline visualization that the *h-index* cluster is new and growing fast. The new cluster includes not only a highly cited article, but also with a strong surge of citations. Hirsch_2005, the original article that introduces the h-index, has the strongest citation burst (See Figure 11).

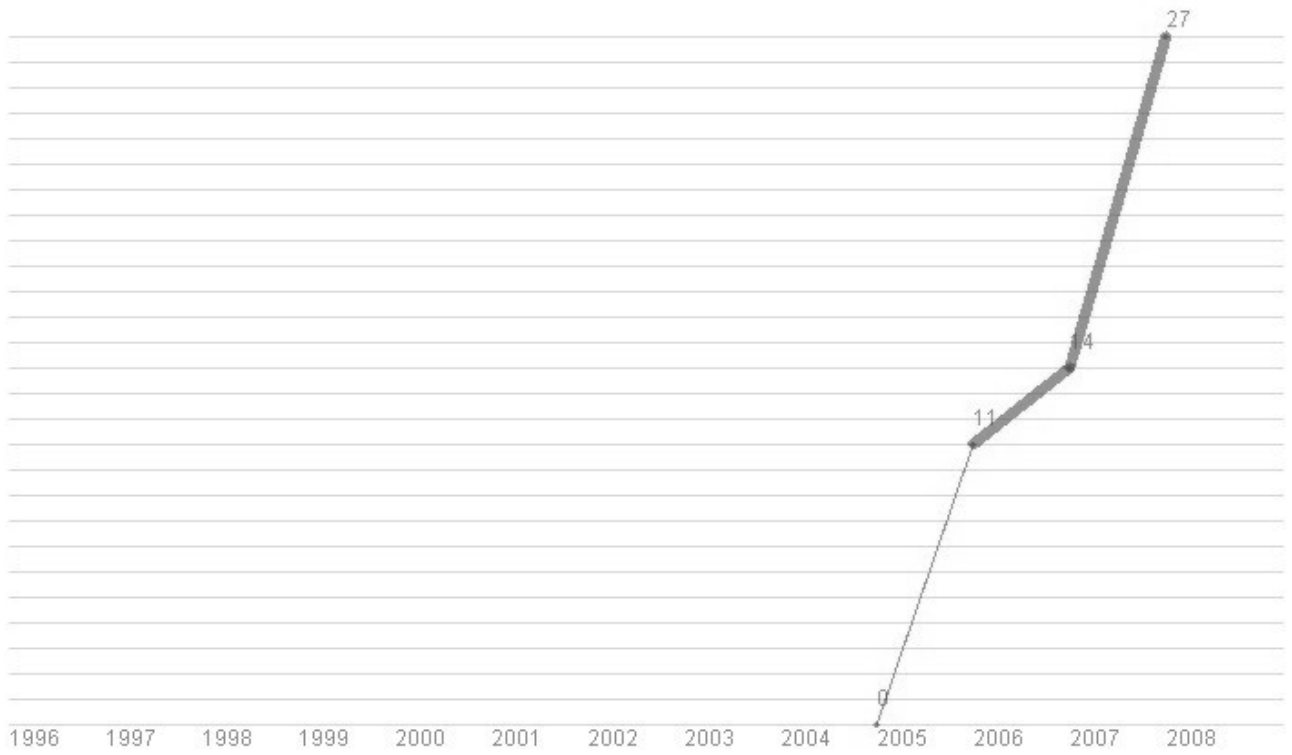

**Figure 11. The burst of citations to Hirsch_2005.**

If we start from the top of the timeline visualization and move down line by line, we can see many representative references in these clusters. For example, *Film Archive* (#11) is a relatively new cluster, containing Jansen_2000 on searching for multimedia on the web as a major reference. Similarly, the most cited reference in *information retrieval* (#13) is Salton's book. Further down the timeline list is the *citation behavior* cluster (#35), which features Garfield_1979 prominently. Many co-citation links join *citation behavior* and *academic web*. Some long-range co-citation links connect the *h-index* cluster and other clusters such as the *academic web* cluster and the *power law* cluster.

## Discussion

We have introduced a multiple-perspective co-citation procedure that consistently supports both ACA and DCA. We have demonstrated how an integration of spectral clustering, automatic cluster labeling, and automatic multiple-abstract summarization can provide informative and insightful cues for analysts to make sense and interpret the nature of co-citation clusters and their interrelationships.

### *Major Findings*

First, the 5-year comparative ACA shows that, at the highest level, both ACA studies identified compatible specialties such as *document retrieval*, *information behavior*, and *webometrics*. On the other hand, spectral clustering identified some distinct clusters that belong to the same factor-induced specialties. For example, the *network diagram* (#4) and *document space* (#8) clusters correspond to the *mapping of science* specialty in Zhao and Strotmann. The study also reveals that human experts tend to choose broader-term labels, equivalent to the level of index terms, whereas algorithmically chosen terms from titles or abstracts tend to be more specific and limited to terms actually used by the authors.

The 13-year progressive ACA (1996-2008) of 633 cited authors in 40 clusters resulted in a clearer global structure than the 5-year single-slice ACA. We have shown that cluster membership and citer-focused labeling provide complementary information to form a comprehensive image of the field.

Finally, the progressive DCA of 655 cited references detects a distinct and fast growing cluster – the *h index* cluster, which is absent from the progressive ACA. The *h index* cluster emerged in 2005. While we do not expect to find it in the comparative ACA of 2001-2005, it is interesting that the 1996-2008 ACA did not detect it. Therefore, we recommend that both ACA and DCA should be conducted whenever possible to maximize the coverage of the intellectual landscape of a subject domain.

### *Limitations and Future Work*

The information science field in our study was defined in terms of the 12 library and information science journals. Although the selection was adopted by several previous studies, the validity of the 12 journals needs to be established in a broader context, for example, by using journal co-citation analysis and subject category analysis. Newer journals such as *Information Retrieval* and *Journal of Informetrics* should be added in future studies and some of the journals may need to be replaced. Furthermore, we exclusively relied on the Web of Science as the single data source. Other sources such as Scopus and Google Scholar would provide additional and unique insights.

We found the comparison with the study of Zhao and Strotmann very valuable. It offered us an opportunity to compare the analysis conducted by human experts to the interpretation cues provided by our automatic labeling and summarization methods. In a long run, we strongly recommend cross-study comparisons so that one can establish a comparable platform to accommodate findings made by different researchers. In part due to the already lengthy article, we decided not to include a comparative DCA study. We focused on co-citation networks in this article. Alternative studies such as bibliographic coupling of journals, authors, documents, and subject categories are also useful methods. Furthermore, we did not include an investigation of direct citation networks in our study. Some recent work in this area can be found in (Garfield, 2004; Morris & Van der Veer Martens, 2008). We recommend that these alternative studies should be considered and compared thoroughly in the future work.

The quality of cluster labeling in general depends on the variety, breadth, and depth of the set of candidate terms. We have extracted candidate cluster labels from citing articles' titles and abstracts. It is also desirable to compare these labels with candidate terms from cited references in the DCA or cited authors' publications in ACA. However, such data is not readily available due to the fact that the CR field in the ISI data format does not include the title of a cited reference. Furthermore, a cited reference may not even be a source record in the entire collection of the Web of Science. In contrast, human analysts may choose the most appropriate label terms from a much wider range of sources beyond the terms found in the immediate dataset, although this could be a double-edged sword. On the one hand, human analysts are free from the limitations of a specific data source. On the other hand, they may need to deal with a potentially much larger search space, which can be a daunting task, especially for those who do not have an encyclopedic knowledge of the subject domain. Utilizing external information sources such as the Wikipedia and the World-Wide Web is a promising direction to resolve the problems due to the limited term space problem. An interesting approach was reported recently in (Carmel et al., 2009).

Sentence selection algorithms in this study were applied to abstracts. Ideally, we expect an improved performance with full text of scientific papers. This is an avenue for further research. A thorough evaluation of the merits of individual labeling approaches is an area in which further research is needed. It is rare to see a citation analysis being evaluated with built-in gold standards in a dataset. Efforts should be made to build a dataset and judgments of domain experts on specialties and the nature of various cluster groupings. Researchers would greatly benefit from such data repositories and benchmark data in order to better validate new techniques and individual interpretations and conclusions.

Spectral clustering for the purpose of network decomposition is exclusive in nature although in reality it is often sensible to allow overlapping clusters due to multiple roles individual entities may play. Spectral clustering of co-citation networks tends to generate distinct clusters with high-precision, whereas human experts tend to aggregate entities into broadly defined clusters. Future work should focus on how to compromise exclusive clusters and how to further aggregate distinct but tightly coupled clusters into super-clusters at a comparable level of abstraction to human experts.

Although some cluster labels make good sense, some labels are still puzzling and some members of clusters may not be as intuitive as others. Some of the labels appear to be strongly biased by particular citing articles, especially when the size of a cluster is relatively small. Algorithmically generated cluster labels are limited to deal with clusters that have multiple aspects formed by a diverse range of citing papers. Clusters with low mean silhouette values tend to be subject to such limitations more than high silhouette clusters. On a positive note, metrics such as modularity and silhouette provide useful indicators of uncertainty that analysts should take into account when interpreting the nature of clusters. We have been looking for a labeling algorithm that is consistently better than others. Since we do not have datasets with gold standards, this cannot be validated systematically except by making comparisons across the 9 sets of candidate labels.

A few more fundamental questions need to be thoroughly addressed. If labels selected from citers differ from those from citees, how do we reconcile the difference? How do we make sense of the citer-citee duality? One of the fundamental assumptions for co-citation analysis is that co-citation clusters do represent something substantial as real as part of the reality although they might be otherwise invisible. Given that some co-citation clusters appear to be biased by the citation behavior of particular publications, it may become necessary to re-examine the assumption, especially whether co-citation clusters represent something that is truly integral to the scientific community as a whole.

## Conclusions

In conclusion, the new co-citation analysis procedure has the following advantages over the traditional one:

1. It can be consistently used for both DCA and ACA.
2. It uses more flexible and efficient spectral clustering to identify co-citation clusters.

3. It characterizes clusters with candidate labels selected by multiple ranking algorithms from the citers of these clusters and reveals the nature of a cluster in terms of how it has been cited.
4. It provides metrics such as modularity and silhouette as quality indicators of clustering to aid interpretation tasks.
5. It provides integrated and interactive visualizations for exploratory analysis.

These features enhance the interpretability and accountability of co-citation analysis. Modularity and silhouette metrics provide useful quality indicators of clustering and network decomposition. This is a valuable addition to the traditional methods such as estimating the strength of a membership based on factor loading. Multiple channels of candidate labels selected from multiple sources have confirmed that clustering labeling is indeed a complex phenomenon that requires multiple perspectives. Our new procedure provides a range of options for analysts to choose cluster labels. In return, the labeling tasks have shed new insights into the behavior of classic information retrieval models and term weighting techniques. Sentence selection was based on weighting functions. In the future we aim to explore the impact of adding linguistic features to these functions such as the argumentative roles of sentences, namely introduction, method, results, conclusion, and novelty, in order to enhance the sentence selection functions. The integration of these techniques in a unifying framework will enable analysts, researchers, and students to investigate and understand the dynamic interrelationship between an intellectual base and an inspired research front. Multiple perspective approaches also provide a cross-validation basis for evaluation and comparison.

The ACA and DCA studies have updated the intellectual landscape formed by a series of previous studies, notably the ACA by White and McCain (1998) and Zhao and Strotmann (2008a, 2008b). The earlier landscape characterized by two predominant camps of information retrieval and citation analysis has evolved into more distinct specialties. The deep impact of the World-Wide Web is evident, responsible for the growth in areas such as webometrics and web search (i.e. the interactive information retrieval cluster in both ACA and DCA). The emergence of the h index in the DCA landscape also indicates the evolution of the information science field. The comparative ACA is a valuable exercise that has characterized the commonality and uniqueness of distinct analytical methods of co-citation networks.

In conclusion, the integrative procedure and the analysis of the information science field contribute to co-citation analysis by improved interpretability and accountability. One can also repeatedly and periodically conduct ACA and DCA with simplified and reduced operational costs. We expect that the integrated implementation provides a rich and evolving platform for research, evaluation, and education for the information science community.

## Acknowledgements

This work is supported in part by the National Science Foundation under grant IIS-0612129. We wish to thank Eric SanJuan of the University of Avignon, France, for implementing the *Enertex* sentence ranking algorithm used in this study; Howard White, Drexel University, USA, for his detailed and constructive comments on an earlier draft; Dangzhi Zhao and Andreas Strotmann of the University of Alberta, Canada, for providing detailed factor loading results of their factor analysis for the comparative ACA and for their comments on an earlier draft of the article; and anonymous reviewers for their detailed reviews. Special thanks to Zeyuan Liu and members of the WISELAB at Dalian University of Technology, China, for their valuable feedback based on their extensive use of earlier versions of *CiteSpace*.

## Notes

*CiteSpace* is freely available at http://cluster.cis.drexel.edu/~cchen/citespace/.
Supplementary materials of the study are available at
http://cluster.cis.drexel.edu/~cchen/papers/2009/infosci/ .

## References

Åström, F. (2007). Changes in the LIS research front: Time-sliced cocitation analyses of LIS journal articles, 1990-2004. Journal of the American Society for Information Science and Technology, 58(7), 947-957.

Bar-Ilan, J. (2008). Which h-index? - A comparison of WoS, Scopus and Google Scholar. Scientometrics, 74(2), 257-271.

Batagelj, V., & Mrvar, A. (1998). Pajek - Program for large network analysis. Connections, 21(2), 47-57.

Ben-Hur, A., Elisseeff, A., & Guyon, I. (2002). A stability based method for discovering structure in clustered data. In R.B. Altman, A.K. Dunker, L. Hunter, K. Lauderdale & T.E. Klein (Eds.), Proceedings of Pacific Symposium on Biocomputing (pp. 6-17). World Scientific.

Bonacich, P. (1987). Power and centrality: A family of measures. American Journal of Sociology, 92, 1170-1182.

Brandes, U. (2001). A faster algorithm for betweenness centrality. Journal of Mathematical Sociology, 25(2), 163-177.

Brin, S., & Page, L. (1998). The anatomy of a large-scale hypertextual web search engine. Proceedings of the 7th International World Wide Web Conference.

Carmel, D., Roitman, H., & Zwerdling, N. (2009). Enhancing cluster labeling using wikipedia. Proceedings of the 32nd international ACM SIGIR conference on Research and development in information retrieval (pp. 139-146).

Chen, C. (1999). Visualising semantic spaces and author co-citation networks in digital libraries. Information Processing & Management, 35(3), 401-420.

Chen, C. (2004). Searching for intellectual turning points: Progressive Knowledge Domain Visualization. Proc. Natl. Acad. Sci. USA, 101(suppl), 5303-5310.

Chen, C. (2005). The centrality of pivotal points in the evolution of scientific networks. Proceedings of the International Conference on Intelligent User Interfaces (IUI 2005) (pp. 98-105). ACM Press.

Chen, C. (2006). CiteSpace II: Detecting and visualizing emerging trends and transient patterns in scientific literature. Journal of the American Society for Information Science and Technology, 57(3), 359-377.

Chen, C., Chen, Y., Horowitz, M., Hou, H., Liu, Z., & Pellegrino, D. (2009a). Towards an explanatory and computational theory of scientific discovery. Journal of Informetrics, 3(3), 191-209.

Chen, C., Song, I.Y., Yuan, X.J., & Zhang, J. (2008). The Thematic and Citation Landscape of Data and Knowledge Engineering (1985-2007). Data and Knowledge Engineering, 67(2), 234-259.

Chen, C., Zhang, J., & Vogeley, M.S. (2009b). Mapping the global impact of Sloan Digital Sky Survey. IEEE Intelligent Systems, 24(4), 74-77.

Cronin, B. (1981). Agreement and Divergence on Referencing Practice. Journal of Information Science, 3(1), 27-33.

Deerwester, S., Dumais, S.T., Landauer, T.K., Furnas, G.W., & Harshman, R.A. (1990). Indexing by Latent Semantic Analysis. Journal of the American Society for Information Science, 41(6), 391-407.

Dunning, T. (1993). Accurate methods for the statistics of surprise and coincidence. Computational Linguistics, 19(1), 61-74.

Fernandez, S., SanJuan, E., & Torres-Moreno, J.M. (2007). Textual energy of associative memories: Performants applications of Enertex algorithm in text summarization and topic segmentation. Proceedings of the 6th Mexican International Conference on Artificial Intelligence (MICAI'07) (pp. 861-871). Springer.

Fiszman, M., Demner-Fushman, D., Kilicoglu, H., & Rindflesch, T.C. (2009). Automatic summarization of MEDLINE citations for evidence-based medical treatment: A topic-oriented evaluation. Journal of Biomedical Informatics, 42, 801-813.

Freeman, L.C. (1977). A set of measuring centrality based on betweenness. Sociometry, 40, 35-41.

Garfield, E. (1979). Citation Indexing: Its Theory and Applications in Science, Technology, and Humanities. New York: John Wiley.

Garfield, E. (2004). Historiographic mapping of knowledge domains literature. Journal of Information Science, 30(2), 119-145.

Ibekwe-SanJuan, F. (2009). Information Science in the web era: A term-based approach to domain mapping. Proceedings of the Annual Meeting of the American Society for Information Science & Technology (ASIS&T 2009).

Jaccard, P. (1901). Étude comparative de la distribution florale dans une portion des Alpes et des Jura. Bulletin del la Société Vaudoise des Sciences Naturelles, 37, 547-579.

Janssens, F., Leta, J., Glanzel, W., & De Moor, B. (2006). Towards mapping library and information science. Information Processing & Management, 42(6), 1614-1642.

Kiss, C., & Bichler, M. (2008). Identification of influencers: Measuring influence in customer networks. Decision Support Systems, 46(1), 233-253.

Klavans, R., Persson, O., & Boyack, K.W. (2009). Coco at the Copacabana: Introducing co-cited author pair co-citation. In B. Larsen & J. Leta (Eds.), Proceedings of the 12th International Conference on Scientometrics and Informetrics (ISSI 2009) (pp. 265-269). BIREME/PAHO/WHO and Federal University of Rio de Janeiro.

Kleinberg, J. (2002). Bursty and hierarchical structure in streams. Proceedings of Proceedings of the 8th ACM SIGKDD International Conference on Knowledge Discovery and Data Mining (pp. 91-101). ACM Press.

Kumar, R., Novak, J., Raghavan, P., & Tomkins, A. (2003). On the bursty evolution of blogspace. Proceedings of WWW2003 (pp. 568-576). ACM.

Lane, J. (2009). Assessing the impact of science funding. Science, 324(5932), 1273-1275.

Leydesdorff, L. (2005). Similarity measures, author cocitation analysis, and information theory. Journal of the American Society for Information Science and Technology, 56(7), 769-772.

Luxburg, U.v. (2006). A tutorial on spectral clustering, from http://www.kyb.mpg.de/publications/attachments/Luxburg06_TR_%5B0%5D.pdf

Luxburg, U.v., Bousquet, O., & Belkin, M. (2009). Limits of spectral clustering, from <http://kyb.mpg.de/publications/pdfs/pdf2775.pdf>

Meho, L.I., & Yang, K. (2007). Impact of data sources on citation counts and rankings of LIS faculty: Web of science versus scopus and google scholar. Journal of the American Society for Information Science and Technology, 58(13), 2105-2125.

Mihalcea, R., & Tarau, P. (2004). TextRank: Bringing Order into Texts. In L. Dekang & W. Dekai (Eds.), Proceedings of the Conference on Empirical Methods in Natural Language Processing (EMNLP 2004) (pp. 8-15). Association for Computational Linguistics.

Morris, S., DeYong, C., Wu, Z., Salman, S., & Yemenu, D. (2002). DIVA: a visualization system for exploring document databases for technology forecasting. Computers and Industrial Engineering, 43(4), 841-862.

Morris, S.A., & Van der Veer Martens, B. (2008). Mapping research specialties. Annual Review of Information Science and Technology, 42, 213-295.

Morris, S.A., Yen, G., Wu, Z., & Asnake, B. (2003). Time line visualization of research fronts. Journal of the American Society for Information Science and Technology, 54(5), 413-422.

Newman, M.E.J. (2006). Modularity and community structure in networks. PNAS, 103(23), 8577-8582.

Ng, A.Y., Jordan, M.I., & Weiss, Y. (2002). On spectral clustering: Analysis and an algorithm. Advanced in Neural Information Processing Systems, 14(2), 849-856.

Persson, O. (1994). The intellectual base and research fronts of JASIS 1986-1990. Journal of the American Society for Information Science, 45(1), 31-38.

Radev, D., & Erkan, G. (2004). Lexrank: Graph-based centrality as salience in text summarization. Journal of Artificial Intelligence Research, 22, 457-480.

Rousseeuw, P.J. (1987). Silhouettes: A graphical aid to the interpretation and validation of cluster analysis. Journal of Computational and Applied Mathematics, 20, 53-65.
Salton, G., Yang, C.S., & Wong, A. (1975). A Vector Space Model for Information Retrieval. Communications of the ACM, 18(11), 613-620.
Schmid, H. (1999). Improvements in part-of-speech tagging with an application to German. In S. Armstrong, K. Church, P. Isabelle, S. Manzi, E. Tzoukermann & D. Yarowsky (Eds.), Natural Language Processing Using Very Large Corpora (pp. 13-26). Dordrecht: Kluwer Academic Publishers.
Schneider, J.W. (Year). Concept symbols revisited: Naming clusters by parsing and filtering of noun phrases from citation contexts of concept symbols. In Proceedings of (pp. 573-593. Springer.
Schneider, J.W. (2009). Mapping of cross-reference activity between journals by use of multidimensional unfolding: Implications for mapping studies. In B. Larsen & J. Leta (Eds.), Proceedings of 12th International Conference on Scientometrics and Informetrics (ISSI 2009) (pp. 443-454). BIREME/PAHO/WHO and Federal University of Rio de Janeiro.
Schneider, J.W., Larsen, B., & Ingwersen, P. (2009). A comparative study of first and all-author co-citation counting, and two different matrix generation approaches applied for author co-citation analyses. Scientometrics, 80(1), 103-130.
Shi, J., & Malik, J. (2000). Normalized Cuts and Image Segmentation. IEEE Transactions on Pattern Analysis and Machine Intelligence, 22(8), 888-905.
Shibata, N., Kajikawa, Y., Taked, Y., & Matsushima, K. (2008). Detecting emerging research fronts based on topological measures in citation networks of scientific publications. Technovation, 28(11), 758-775.
Small, H. (1973). Co-citation in the scientific literature: A new measure of the relationship between two documents. Journal of the American Society for Information Science, 24, 265-269.
Small, H. (1978). Cited documents as concept symbols. Social Studies of Science, 8(3), 327-340.
Small, H. (1980). Co-Citation context analysis and the structure of paradigms. Journal of Documentation, 36(3), 183-196.
Small, H. (1986). The synthesis of specialty narratives from co-citation clusters. Journal of the American Society for Information Science, 37(3), 97-110.
Small, H. (2003). Paradigms, citations, and maps of science: A personal history. Journal of the American Society for Information Science and Technology, 54(5), 394-399.
Small, H., & Greenlee, E. (1986). Collagen Research in the 1970s. Scientometrics, 10(1-2), 95-117.
Small, H., & Sweeney, E. (1985). Clustering the Science Citation Index Using Co-Citations .1. a Comparison of Methods. Scientometrics, 7(3-6), 391-409.
Small, H., Sweeney, E., & Greenlee, E. (1985). Clustering the Science Citation Index Using Co-Citations .2. Mapping Science. Scientometrics, 8(5-6), 321-340.
Sparck Jones, K. (1999). Automatic Summarizing: Factors and Directions. In I. Mani & M.T. Maybury (Eds.), Advances in Automatic Text Summarization (pp. 2-12). Cambridge, MA: MIT Press.
Tabah, A.N. (1999). Literature dynamics: Studies on growth, diffusion, and epidemics. Annual Review of Information Science and Technology, 34, 249-286.
Teufel, S., & Moens, M. (2002). Summarizing scientific articles: Experiments with relevance and rhetorical status. Computational Linguistics, 28(4), 409-445.
White, H.D. (2007a). Combining bibliometrics, information retrieval, and relevance theory, Part 1: First examples of a synthesis. Journal of the American Society for Information Science and Technology, 58(4), 536-559.
White, H.D. (2007b). Combining bibliometrics, information retrieval, and relevance theory, Part 2: Some implications for information science. Journal of the American Society for Information Science and Technology, 58(4), 583-605.
White, H.D., & Griffith, B.C. (1982). Authors as Markers of Intellectual Space - Co-Citation in Studies of Science, Technology and Society. Journal of Documentation, 38(4), 255-272.

White, H.D., & McCain, K.W. (1998). Visualizing a discipline: An author co-citation analysis of information science, 1972-1995. Journal of the American Society for Information Science, 49(4), 327-355.

Witten, I.H., & Frank, E. (1999). Data Mining: Practical Machine Learning Tools and Techniques with Java Implementations. San Francisco, CA: Morgan Kaufmann.

Zhao, D.Z., & Strotmann, A. (2008a). Evolution of Research Activities and Intellectual Influences in Information Science 1996-2005: Introducing Author Bibliographic-Coupling Analysis. Journal of the American Society for Information Science and Technology, 59(13), 2070-2086.

Zhao, D.Z., & Strotmann, A. (2008b). Information science during the first decade of the web: an enriched author cocitation analysis. Journal of the American Society for Information Science and Technology, 59(6), 916-937.

Zins, C. (2007a). Classification schemes of information science: Twenty-eight scholars map the field. Journal of the American Society for Information Science and Technology, 58(5), 645-672.

Zins, C. (2007b). Conceptions of information science. Journal of the American Society for Information Science and Technology, 58(3), 335-350.

Zins, C. (2007c). Conceptual approaches for defining data, information, and knowledge. Journal of the American Society for Information Science and Technology, 58(4), 479-493.

Zins, C. (2007d). Knowledge map of information science. Journal of the American Society for Information Science and Technology, 58(4), 526-535.

Zuccala, A. (2006). Author cocitation analysis is to intellectual structure as web colink analysis is to ... ? Journal of the American Society for Information Science and Technology, 57(11), 1487-1502.